\documentclass[12pt]{article}
\usepackage[utf8]{inputenc}
\usepackage[margin=1in]{geometry}

\usepackage{graphbox}
\usepackage{pbox}
\usepackage{enumitem}
\usepackage{parskip}
\usepackage{booktabs}
\usepackage{caption}
\usepackage{float}
\usepackage{authblk}

\usepackage{amsmath,amsfonts,amssymb}
\usepackage{esdiff,commath}
\usepackage{siunitx}

\usepackage{hyperref}
\usepackage[noabbrev]{cleveref}

\usepackage{natbib}
\setcitestyle{round,semicolon}

\DeclareSIUnit\cells{cell}
\begin{document}

\title{Macrophage anti-inflammatory behaviour in a
    multiphase model of atherosclerotic plaque development}
\author[1]{Ishraq U.~Ahmed}
\affil{School of Mathematics and Statistics, University of Sydney}
\author[2]{Helen M.~Byrne}
\affil{Wolfson Centre for Mathematical Biology,
    Mathematical Institute, University of Oxford}
\author[1]{Mary R.~Myerscough}
\maketitle

\begin{abstract}
    
Atherosclerosis is an inflammatory disease characterised by the formation of plaques,
which are deposits of lipids and cholesterol-laden macrophages
that form in the artery wall.
The inflammation is often non-resolving, due in large part to
changes in normal macrophage anti-inflammatory behaviour
that are induced by the toxic plaque microenvironment.
These changes include higher death rates,
defective efferocytic uptake of dead cells,
and reduced rates of emigration.
We develop a free boundary multiphase model for early atherosclerotic plaques,
and we use it to investigate the effects of impaired macrophage anti-inflammatory behaviour
on plaque structure and growth.
We find that high rates of cell death relative to efferocytic uptake
results in a plaque populated mostly by dead cells.
We also find that emigration can potentially slow or halt plaque growth
by allowing material to exit the plaque,
but this is contingent on the availability of live macrophage foam cells in the deep plaque.
Finally, we introduce an additional bead species
to model macrophage tagging via microspheres,
and we use the extended model to explore how
high rates of cell death and low rates of efferocytosis and emigration
prevent the clearance of macrophages from the plaque.
    
\end{abstract}

\section{Introduction}

Atherosclerosis is a chronic inflammatory disorder characterised by
the retention of lipoproteins and
cholesterol-laden macrophages in the artery wall
\citep{libby2019atherosclerosis,tabas2015recent}.
Certain macrophage behaviours are vital in enabling inflammation to resolve
in both plaques and other inflammatory contexts.
These behaviours include the emigration of macrophages
from the site of inflammation
and the clearance of dead cells
\citep{tabas2010macrophage,llodra2004emigration}.
Macrophages in advanced plaques, however, display an impaired ability to
carry out normal anti-inflammatory functions
\citep{tabas2010macrophage,randolph2008emigration}.
Understanding how these impaired behaviours may
promote macrophage retention in early plaques
is therefore important to identify the conditions
that lead to advanced plaque development.
In this paper, we develop a partial differential equation model
for an atherosclerotic plaque based on a multiphase framework,
and we use this to investigate how macrophage emigration and dead cell clearance
influence the growth and composition of the early plaque.

Atherosclerotic lesions are initiated when
cholesterol-carrying low density lipoproteins (LDL)
from the bloodstream are deposited in the intima
\citep{libby2019atherosclerosis,falk2006pathogenesis}.
The intima is an initially thin layer that separates the endothelium
(a monolayer of endothelial cells lining the interior of the artery)
from the media (a thicker layer comprising muscle cells and collagen).
The endothelium becomes more permeable to particles such as LDL
at sites of endothelial dysfunction,
which occur due to disrupted blood shear flow or endothelial cell damage
\citep{gimbrone2013vascular}.
LDL particles that have penetrated the endothelium
and accumulated in the intima
undergo oxidation and other forms of chemical modification
to produce modified LDL (modLDL)
\citep{madamanchi2005oxidative,yoshida2010mechanisms}.
Intimal modLDL is a potent immune trigger,
and stimulates the expression of adhesion molecules by endothelial cells.
These adhesion molecules bind to and capture
monocytes circulating in the bloodstream,
which then transmigrate through the endothelium into the intima
\citep{blankenberg2003adhesion,bobryshev2006monocyte}.
Once in the intima, monocytes differentiate into macrophages,
which then take up modLDL via phagocytosis
\citep{bobryshev2006monocyte,moore2013macrophages}.
The resulting cholesterol-engorged macrophages are
referred to as macrophage foam cells.
Early atherosclerotic lesions consist largely of
modLDL, macrophage foam cells, and debris from dead cells.

In atherosclerosis and other forms of inflammation,
macrophages undergo a form of programmed cell death called apoptosis
\citep{cohen2013extrinsic,van2012apoptotic}.
Apoptosis is normal even in healthy tissue,
and apoptotic cells express find-me and eat-me signals
to encourage their clearance by live macrophages via efferocytosis.
In chronic inflammatory conditions such as advanced atherosclerosis,
the accumulation of apoptotic material
will often overwhelm macrophages' efferocytic capacity.
This is especially likely in cases where
rates of macrophage death are elevated due to
high levels of ingested cytotoxic material (such as cholesterol),
or where efferocytosis itself becomes defective.
Both of these are observed in atherosclerotic plaques
\citep{yurdagul2018mechanisms,schrijvers2005phagocytosis}.
Uncleared apoptotic cells will eventually undergo
an uncontrolled form of cell death called necrosis.
Necrotic macrophages are highly problematic
due to their lower production of ``find-me'' and ``eat-me'' signals.
Necrotic material is also highly inflammatory
and will attract more macrophages
which are themselves likely to undergo necrosis,
thereby perpetuating the cycle.
A major characteristic of advanced plaques
is the presence of a large necrotic core,
consisting of lipids and debris released by necrotic cells
\citep{sakakura2013pathophysiology}.
The inefficient efferocytic clearance of apoptotic material
is therefore an important precursor to
the development of a vulnerable plaque state
that is more likely to rupture.

The emigration of macrophages from sites of inflammation is a critical part
of inflammation resolution in other inflammatory conditions
\citep{lawrence2007chronic,bellingan1996vivo},
and is believed to play an important part in atherosclerosis as well,
although its mechanisms are not yet well understood.
Experimental studies have measured the expression of receptors such as CCR7
that are known to be involved in macrophage emigration
in other inflammatory conditions,
and find that higher expression of these receptors
is correlated with reduced plaque size
\citep{trogan2002laser,finney2017integrin}.
Several studies suggest that emigration most likely happens
via migration into lymphatic vessels that connect to the outer artery.
One such study found that after plaque regression was induced,
macrophages that were originally present in the plaque
were found in the lymph nodes \citep{llodra2004emigration}.
Another study involving fluorescent bead-tagged macrophages
found that the distance between tagged macrophages and the elastic lamina
(a membrane separating the intima from the outer muscle cell layers)
decreased during plaque regression,
suggesting that macrophages are exiting through the outer artery
instead of transmigrating across the endothelium \citep{williams2018limited}.

Mathematical modelling of the inflammatory response during atherosclerosis
is an area of growing interest \citep{parton2016computational}.
Much of this work uses a continuum approach to investigate
the activity of monocytes and macrophages in the vessel wall,
and their interactions with lipoproteins.
Previous work in our group has considered
the accumulation and efferocytic clearance of apoptotic cells
using both ODE models \citep{lui2021modelling}
and non-spatial lipid-structured PDE models
for macrophage populations in plaques \citep{ford2019lipid}.
An alternative model by \citet{fok2012growth} considers macrophages and dead cells
in a reaction-diffusion model for plaque development.
No other spatial models exist to date
that explicitly consider both cell death and efferocytosis.
Another shortcoming of many spatial models is
their use of a fixed spatial domain to model the intima,
where the domain remains fixed in size
rather than expanding as the plaque develops.
Some models employ a moving boundary to allow the artery wall to expand
\citep{friedman2015mathematical,thon2018multiphysics}.
In particular, one model by \citet{friedman2015mathematical} 
employs a multiphase framework to capture the effects of cell crowding.
However, none of these models incorporate macrophage emigration
in a way that respects local mass conservation,
although some fixed domain models incorporate an unbalanced sink term
for macrophage removal due to death or emigration
\citep{fok2012growth}.

In this paper, we present a free boundary multiphase model
for early plaque growth that focuses on
the interplay between foam cell death, efferocytosis, and emigration.
We formulate the model in \Cref{sec:2}.
In \Cref{sec:3}, we investigate how the relative rates
of foam cell death and efferocytic uptake influence plaque composition.
In \Cref{sec:4}, we show how macrophage emigration
can slow or halt plaque growth,
and how its effectiveness depends on the availability
of live foam cells in the deep plaque.
In \Cref{sec:5}, we extend the model to include macrophage tagging,
and we use the extended model to investigate
macrophage retention in plaques.
We conclude in \Cref{sec:discussion} with a discussion of the results
in the context of existing plaque models and biomedical studies,
and some suggestions for future modelling work.

\section{Model formulation}
\label{sec:2}

\begin{figure}
    \centering
    \includegraphics[align=t,width=0.8\textwidth]
    {{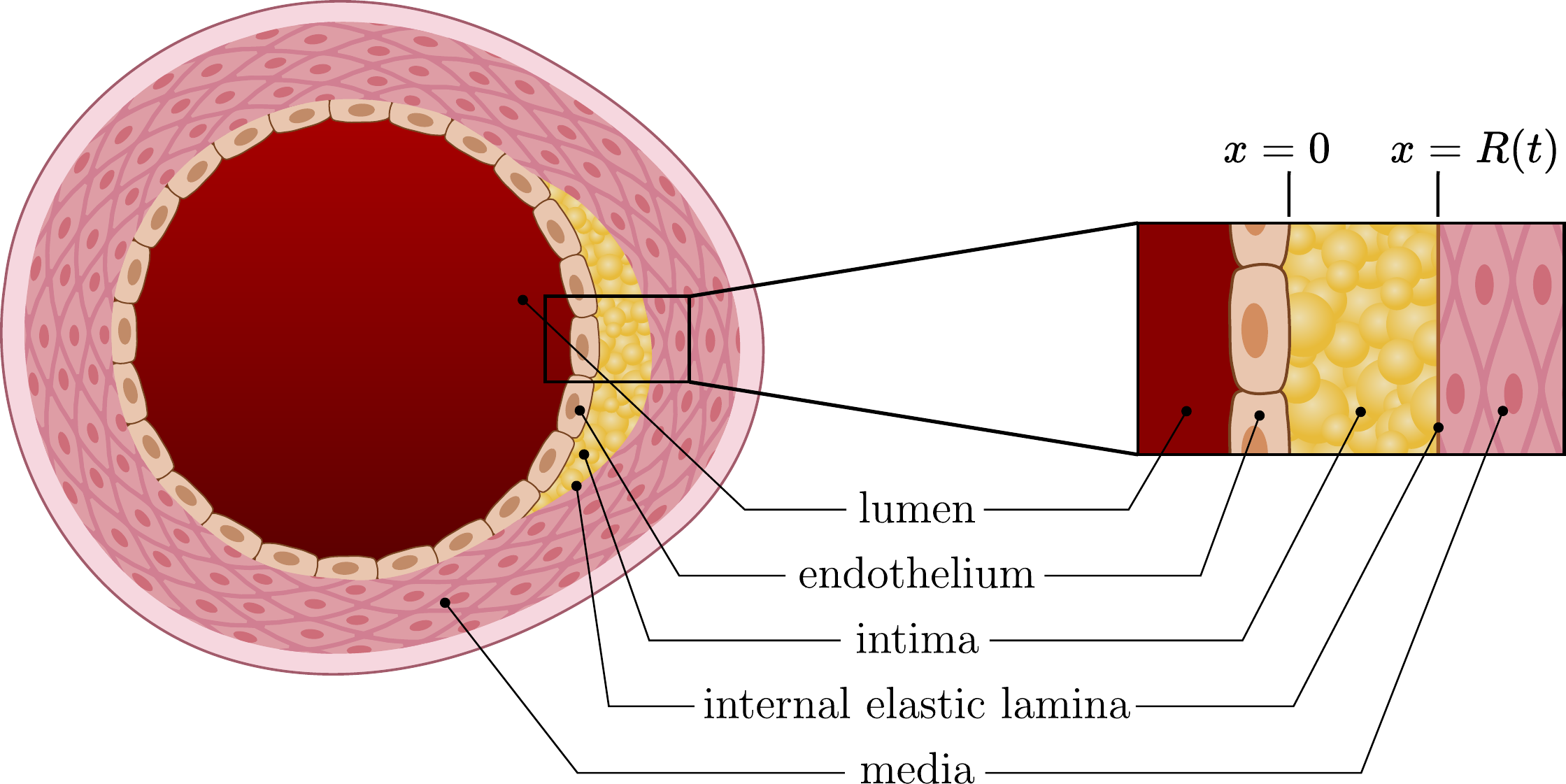}} \\ \vspace{6pt}
    \caption{A 2-dimensional cross section of an artery,
        with a 1-dimensional radial cross section
        taken through the plaque for the problem domain. }
    \label{fig:2-domain}
\end{figure}

In this section we present a general model that describes
the growth of an early stage atherosclerotic plaque.
We model the intima as a 1-dimensional domain
on the interval $[0,\tilde{R}(\tilde{t})]$,
which represents a radial cross-section through the artery wall.
The $\tilde{x} = 0$ boundary corresponds to the endothelium,
and the $\tilde{x} = \tilde{R}(\tilde{t})$ boundary corresponds to
the elastic lamina that separates the intima from the media
(see \Cref{fig:2-domain}).
We allow the $\tilde{x} = \tilde{R}(\tilde{t})$ boundary
to vary with time since early plaques cause the artery
to undergo compensatory enlargement.
During compensatory enlargement,
the artery expands outwards in a way that preserves lumenal area
\citep{korshunov2007vascular}.
We also assume that the diameter of the artery is much larger
than the width of the intima in early stage plaques
\citep{bonithon1996relation},
so that Cartesian coordinates are a good approximation
to radial coordinates.
We notate all quantities with tildes for the full dimensional model,
and then remove tildes for the nondimensional model
in \Cref{subsec:2-nondim}.

We assume that the plaque comprises three phases,
each measured in 
cells or equivalent cell volumes per unit length.
These phases are macrophage foam cells $\tilde{f}(\tilde{x},\tilde{t})$,
modified LDL (or modLDL) $\tilde{l}(\tilde{x},\tilde{t})$,
and dead cellular material $\tilde{c}(\tilde{x},\tilde{t})$.
For simplicity, we let $\tilde{f}(\tilde{x},\tilde{t})$ encompass
all monocyte-derived cells,
including M1 and M2 macrophages and dendritic cells for instance
\citep{murray2011protective},
ignoring differences in phenotypic expression.
Similarly, we ignore the distinction between
the various forms of chemically modified LDL,
treating them as a single modLDL species
$\tilde{l}(\tilde{x},\tilde{t})$.
For cell death, we ignore the distinction
between apoptosis and necrosis,
and use a generic cell death term to encompass both processes.

The phases obey the following continuity equations:
\begin{align}
\frac{\partial\tilde{f}}{\partial\tilde{t}} &=
    - \frac{\partial}{\partial\tilde{x}}
    ( \tilde{J}_f + \tilde{v} \tilde{f} ) + \tilde{s}_f \,, 
    \label{eq:2-conteqf} \\
\frac{\partial\tilde{l}}{\partial\tilde{t}} &=
    - \frac{\partial}{\partial\tilde{x}}
    ( \tilde{J}_l + \tilde{v} \tilde{l} ) + \tilde{s}_l \,,
    \label{eq:2-conteql} \\
\frac{\partial\tilde{c}}{\partial\tilde{t}} &=
    - \frac{\partial}{\partial\tilde{x}}
    ( \tilde{J}_c + \tilde{v} \tilde{c} ) + \tilde{s}_c
    \label{eq:2-conteqc} \,.
\end{align}
In the above,
$\tilde{J}_u$ and $\tilde{s}_u$ denote phase-specific
flux and source terms respectively for the phases $u = f,\,l,\,c$,
and $\tilde{v}$ is a common mixture velocity with which
all phases are advected.
We also assume there are no voids throughout the plaque,
so that the phases obey a no-voids condition
\begin{equation} \label{eq:2-novoids-dim}
\tilde{f} + \tilde{l} + \tilde{c} = N_0
\end{equation}
everywhere,
where $N_0$ denotes the total phase density,
which is constant with time and space.

Foam cells internalise modLDL via phagocytosis,
and dead material via efferocytosis.
We model both processes with a simple mass-action term.
Foam cells also undergo cell death,
which we model using a linear death term.
This gives the following expressions for the source terms
in \cref{eq:2-conteqf,eq:2-conteql,eq:2-conteqc}:
\begin{alignat}{4}
\tilde{s}_f &= \,&+ \tilde{\mu}_p \tilde{f} \tilde{l} &+
        \tilde{\mu}_e \tilde{f} \tilde{c} - \tilde{\mu}_a \tilde{f} &\,, \\
\tilde{s}_l &= \,&- \tilde{\mu}_p \tilde{f} \tilde{l} & &\,, \\
\tilde{s}_c &= &&- \tilde{\mu}_e \tilde{f} \tilde{c} +
        \tilde{\mu}_a \tilde{f} &\,,
\end{alignat}
where $\tilde{\mu}_a$ is the macrophage foam cell death rate,
$\tilde{\mu}_p$ is the rate of phagocytosis per foam cell
per unit of available modLDL,
and $\tilde{\mu}_c$ is the rate of efferocytosis per foam cell
per unit of available dead material.
Since $\tilde{s}_f + \tilde{s}_l + \tilde{s}_c = 0$,
the model is locally mass conservative.

We assume that foam cells undergo undirected random motion \citep{owen1997pattern}
and directed chemotactic motion towards modLDL.
The latter is a simplification of the full process
whereby modLDL stimulates the production of chemoattractants.
Foam cells also undergo chemotactic motion towards dead material,
in response to find-me signals expressed by the dying cells
\citep{kojima2017role}.
ModLDL and dead material undergo diffusive motion.
For simplicity, we assume constant diffusion coefficients for all phases,
and linear chemotaxis with regards to modLDL and dead material for foam cells.
This gives the flux terms
\begin{align}
\tilde{J}_f &= - \tilde{D}_f \frac{\partial\tilde{f}}{\partial\tilde{x}} +
    \tilde{\chi}_l \tilde{f} \frac{\partial\tilde{l}}{\partial\tilde{x}} +
    \tilde{\chi}_c \tilde{f} \frac{\partial\tilde{c}}{\partial\tilde{x}} \,,\\
\tilde{J}_l &= - \tilde{D}_l \frac{\partial\tilde{l}}{\partial\tilde{x}} \,, \\
\tilde{J}_c &= - \tilde{D}_c \frac{\partial\tilde{c}}{\partial\tilde{x}} \,,
\end{align}
where $\tilde{D}_f$ is the foam cell random motility coefficient,
$\tilde{D}_l$ and $\tilde{D}_c$ are diffusion coefficients
for modLDL and dead material respectively,
and $\tilde{\chi}_l$ and $\tilde{\chi}_c$ are
chemotactic coefficients for foam cells
in response to modLDL and dead material respectively.

The endothelium permits a net influx of LDL
carried by native LDL particles,
which we assume is constant over plaque growth timescales
of days or weeks.
We also assume that typical timescales for chemical modification of LDL
are much faster than plaque growth timescales \citep{cobbold2002lipoprotein}.
Under this assumption, native LDL becomes inflammatory modLDL
immediately after entering the intima,
so we model native LDL deposition as an boundary modLDL influx.
Monocyte recruitment requires
adhesion molecules expressed by endothelial cells
and chemokines produced in the intima.
These are expressed in response to the presence of modified LDL.
For simplicity, we assume monocyte recruitment is proportional
to the amount of modLDL at the endothelium.
We also assume that monocyte differentiation timescales
are much faster than those for plaque growth \citep{blanchard1991production},
and so we model monocyte recruitment as a macrophage boundary influx.
Cells do not die until they are already inside the intima,
so the flux of dead material at the endothelial boundary is zero.
This gives the following flux boundary conditions
at $\tilde{x} = 0$:
\begin{align}
( \tilde{J}_f + \tilde{v} \tilde{f} )
    &= \tilde{\sigma}_f \tilde{l} \,,
    \label{eq:2-bc0f} \\
( \tilde{J}_l + \tilde{v} \tilde{l} )
    &= \tilde{\sigma}_l \,,
    \label{eq:2-bc0l} \\
( \tilde{J}_c + \tilde{v} \tilde{c} ) &= 0 \,,
    \label{eq:2-bc0c} 
\end{align}
where $\tilde{\sigma}_l$ is the rate of LDL deposition,
and $\tilde{\sigma}_f$ is the rate of monocyte recruitment
per unit of modLDL at the endothelium.

At the medial boundary,
we assume zero flux boundary conditions for
the passive modLDL and dead material phases.
We assume foam cells will emigrate into the lymphatic vasculature
at a rate proportional to the quantity of foam cells present.
The corresponding flux boundary conditions
at $\tilde{x} = \tilde{R}$ are given by
\begin{align}
( \tilde{J}_f + \tilde{v} \tilde{f} )
    - \tilde{R}'(\tilde{t}) \tilde{f}
    &= \tilde{\sigma}_e \tilde{f} \,,
    \label{eq:2-bcRf} \\
( \tilde{J}_l + \tilde{v} \tilde{l} )
    - \tilde{R}'(\tilde{t}) \tilde{l} &= 0 \,,
    \label{eq:2-bcRl} \\
( \tilde{J}_c + \tilde{v} \tilde{c} )
    - \tilde{R}'(\tilde{t}) \tilde{c} &= 0 \,, \label{eq:2-bcRc}
\end{align}
where $\sigma_e$ is the average foam cell egress velocity.

For the initial conditions,
we assume the intima is populated with
a small number of live resident macrophages \citep{ley2011monocyte}, so
\begin{align}
\tilde{f}(\tilde{x},0) &= N_0 \,, \\
\tilde{l}(\tilde{x},0) &= 0 \,, \\
\tilde{c}(\tilde{x},0) &= 0 \,, \\
\tilde{R}(0) &= x_S \label{eq:2-icR} \,,
\end{align}
where $x_S$ is the diameter of a typical mouse macrophage.

In order to close the model,
the mixture velocity $\tilde{v}(\tilde{x},\tilde{t})$
and the velocity of the medial boundary $\tilde{R}'(\tilde{t})$
need to be defined, in addition to initial conditions.
Summing over the phase continuity equations
\labelcref{eq:2-conteqf,eq:2-conteql,eq:2-conteqc}
and applying the no-voids condition \labelcref{eq:2-novoids-dim} gives
\begin{alignat}{3} \label{eq:2-conteqtotal}
0 &= - \frac{\partial}{\partial\tilde{x}} \del[3]
    { - \tilde{D}_f \frac{\partial\tilde{f}}{\partial\tilde{x}}
    + \tilde{\chi}_l \tilde{f} \frac{\partial\tilde{l}}{\partial\tilde{x}}
    + \tilde{\chi}_c \tilde{f} \frac{\partial\tilde{c}}{\partial\tilde{x}}
    - \tilde{D}_l \frac{\partial\tilde{l}}{\partial\tilde{x}}
    - \tilde{D}_c \frac{\partial\tilde{c}}{\partial\tilde{x}}
    + N_0 \tilde{v} } \,.
\end{alignat}

Integrating \cref{eq:2-conteqtotal} from $\tilde{x} = 0$ to $\tilde{x}$
and substituting the $\tilde{x} = 0$ boundary conditions
\labelcref{eq:2-bc0f,eq:2-bc0l,eq:2-bc0c}
gives the following expression for the mixture velocity:
\begin{equation}
\tilde{v} = \frac{1}{N_0} \sbr[3]{
    \tilde{\sigma}_l + \tilde{\sigma}_f l \Big\vert_{\tilde{x}=0}
    + \tilde{D}_f \frac{\partial\tilde{f}}{\partial\tilde{x}}
    - \tilde{\chi}_l \tilde{f} \frac{\partial\tilde{l}}{\partial\tilde{x}}
    - \tilde{\chi}_c \tilde{f} \frac{\partial\tilde{c}}{\partial\tilde{x}}
    + \tilde{D}_l \frac{\partial\tilde{l}}{\partial\tilde{x}}
    + \tilde{D}_c \frac{\partial\tilde{c}}{\partial\tilde{x}} } \,.
\end{equation}

The boundary velocity can be similarly obtained by
integrating \cref{eq:2-conteqtotal} from $\tilde{x} = 0$ to $\tilde{R}$
and substituting the $\tilde{x} = 0$ and $\tilde{R}$ boundary conditions
\labelcref{eq:2-bc0f,eq:2-bc0l,eq:2-bc0c,eq:2-bcRf,eq:2-bcRl,eq:2-bcRc},
which yields
\begin{equation}
\frac{\dif\tilde{R}'}{\dif\tilde{t}} = \frac{1}{N_0} \sbr[3]{
    \tilde{\sigma}_l + \tilde{\sigma}_f l \Big\vert_{\tilde{x}=0}
    - \tilde{\sigma}_e f \Big\vert_{\tilde{x}=\tilde{R}} } \,.
    \label{eq:2-dRdt}
\end{equation}

The full dimensional model consists of
\crefrange{eq:2-conteqf}{eq:2-icR} and \labelcref{eq:2-dRdt}.

\subsection{Parameter estimates and model nondimensionalisation}
\label{subsec:2-nondim}

We nondimensionalise the model by applying the rescalings
\begin{equation}
f = \frac{\tilde{f}}{N_0} \,,\quad
l = \frac{\tilde{l}}{N_0} \,,\quad
c = \frac{\tilde{l}}{N_0} \,,\quad
t = \frac{\tilde{t}}{t_S} \,,\quad
x = \frac{\tilde{x}}{x_S} \,,
\end{equation}
where $t_S$ and $x_S$ are some reference timescale and length scale,
and $N_0$ is the total phase density.

We choose $t_S$ to be one week
as experimental studies on plaques in mouse models
typically take place over periods of weeks to months
\citep{williams2018limited,potteaux2011suppressed}.
Lengths are rescaled by the diameter of a typical mouse macrophage.
As observed macrophage sizes vary from under \SI{14}{\micro\metre}
for murine bone marrow-derived macrophages \citep{cannon1992macrophage}
to \SI{20}{\micro\metre} for murine peritoneal macrophages
\citep{nguyen2012il10},
we choose a characteristic size of \SI{16}{\micro\metre}.
Estimating certain parameters requires the conversion
of three-dimensional volume densities
to one-dimensional linear densities.
For the purposes of parameter estimation,
we consider a cylindrical section through the intima
whose circular cross-section is transverse to
the positive $x$ vector 
and has diameter $x_S$,
so that the maximum linear phase density $N_0$
is one cell per unit of $x_S$.

Foam cell motility parameters $\tilde{D}_f$ and $\tilde{\chi}_l$
are estimated using models fitted to data from
chemotaxis assay experiments \citep{owen1997pattern,sozzani1991signal}.
The phagocytosis parameter $\tilde{\mu}_p$ is based on
experimental observations
where macrophages are observed to phagocytose
a significant volume of modLDL after a timescale
of approximately \SI{60}{\minute} \citep{zwaka2001creactive}.
The boundary monocyte recruitment parameter $\tilde{\sigma}_f$ is based on
an experimental study by \citet{jeng1993oxidized}
where monocytes pretreated with modLDL are exposed to shear flow
and are observed to bind to an endothelial cell monolayer
with an area of \SI{0.1452}{\milli\metre\squared} at a rate of
\numrange{140}{200} cells over \SI{30}{\minute}.
The death rate $\tilde{\mu}_a$ is based on experimental observations
of mouse macrophages incubated in acetylated LDL,
where $12\%$ of the initial macrophage population
is observed to undergo apoptosis after $\SI{9}{\hour}$.
Fitting this to a simple exponential decay curve
gives an apoptosis parameter of about $\SI{0.25}{\per\hour}$
before nondimensionalising.

The remaining parameters are based on
order of magnitude estimates.
The modLDL diffusion coefficient $\tilde{D}_l$ is estimated
to be larger than $\tilde{D}_f$ due to the small size of LDL particles
compared to macrophages.
Experimental studies of LDL phagocytosis
\citep{zwaka2001creactive,suits1989phagocytosis}
find that for heavily cholesterol-laden foam cells,
cholesterol droplets comprise a significant proportion
of the total foam cell volume,
and so the range of LDL influx rates is chosen
so that $\tilde{\sigma}_l$ is of a similar order of magnitude
to the monocyte influx $\tilde{\sigma}_f \tilde{l}|_{\tilde{x}=0}$.
The upper range of foam cell egress velocities $\tilde{\sigma}_e$
is chosen so that the medial egress flux

$\tilde{\sigma}_e \tilde{f}|_{\tilde{x}=\tilde{R}}$
is of a similar order of magnitude
to the endothelial LDL and monocyte fluxes
$\tilde{\sigma}_l$ and $\tilde{\sigma}_f \tilde{l}|_{\tilde{x}=0}$.
For efferocytosis, we choose a range of values for $\tilde{\mu}_e$
for which efferocytic uptake rates
are of a similar order of magnitude to apoptosis rates.
Dead material diffusion is assumed to be
slower than live macrophage random motility
due to the lack of active locomotion.
We choose $\tilde{D}_c$ to be lower than $\tilde{D}_f$ within an order of magnitude.
The macrophage chemotactic response to dead material $\tilde{\chi}_c$
is assumed to be lower than that for modLDL,
since the dead material phase includes necrotic material
which expresses few find-me signals.

\Cref{tab:2-param-dim,tab:2-param-nondim} summarise
the model parameter values before and after rescaling.

\begin{table}[h]
\footnotesize \centering

\caption{Dimensional parameter estimates
        (order of magnitude estimates used where
        quantitative biological data was unavailable)}
\label{tab:2-param-dim}

\begin{tabular}{p{0.5cm} p{6cm} p{4.6cm} p{3.2cm}}
    
    \toprule
    \multicolumn{2}{l}{Parameter + Description}
    & Dimensional estimate & Sources \\
    \midrule
    
    $t_S$ & reference time scale
    & \SI{6e5}{\second}
    & --- \\
    
    $x_S$ & reference length scale
    & \SI{1.6e-5}{\metre}
    & \citep{cannon1992macrophage} \\
    
    $N_0$ & maximum phase density
    & $\SI{1}{\cells}/x_S$
    & --- \\
    
    $\tilde{D}_f$ & foam cell random motility coefficient
    & \SIrange{4e-15}{e-14}{\metre\squared\per\second}
    & \citep{owen1997pattern} \\
    
    $\tilde{D}_l$ & modLDL diffusion coefficient
    & $> \tilde{D}_f$
    & order of mag est \\
    
    $\tilde{D}_c$ & dead material diffusion coefficient
    & $< \tilde{D}_f$
    & order of mag est \\
    
    $\tilde{\chi}_l$ & foam cell chemotactic coefficient (modLDL)
    & $\gg \tilde{D}_f/N_0$
    & order of mag est, \citep{owen1997pattern} \\
    
    $\tilde{\chi}_c$ & foam cell chemotactic coefficient (dead material)
    & $\sim \tilde{\chi}_l$
    & order of mag est \\
    
    $\tilde{\mu}_a$ & foam cell death rate
    & $\sim \SI{0.25}{\per\hour}$
    & \citep{yao2000free} \\
    
    $\tilde{\mu}_e$ & efferocytosis rate
    & $\sim \tilde{\mu}_a / N_0$
    & order of mag est, \citep{yao2000free} \\
    
    $\tilde{\mu}_p$ & modLDL phagocytosis rate
    & $\sim (\SI{30}{\minute})^{-1}/N_0$
    & \citep{zwaka2001creactive} \\
    
    $\tilde{\sigma}_f$ & monocyte recruitment rate per unit modLDL
    & $\sim \SI{5e5}{\cells\per\metre\squared\per\second} \cdot x_S^2 /N_0$
    & \citep{jeng1993oxidized} \\
    
    $\tilde{\sigma}_l$ & LDL deposition rate
    & $\sim \tilde{\sigma}_f N_0$
    & order of mag est \\
    
    $\tilde{\sigma}_e$ & foam cell egress velocity
    & $\sim \tilde{\sigma}_f ,\, \tilde{\sigma}_l/N_0$
    & order of mag est \\
    
    \bottomrule
    
\end{tabular}
\end{table}

\begin{table}[h]

\footnotesize \centering

\caption{Rescaled parameters for the nondimensionalised model}
\label{tab:2-param-nondim}

\begin{tabular}{p{2cm} p{3cm} p{4cm}}
    
    \toprule
    Parameter & Rescaling & Nondimensional estimate \\
    \midrule
    
    $D_f$ & $\tilde{D}_f/(x_S^2/t_S)$
    & $20$ \\
    
    $D_l$ & $\tilde{D}_l/(x_S^2/t_S)$
    & $200$ \\
    
    $D_c$ & $\tilde{D}_c/(x_S^2/t_S)$
    & $10$ \\
    
    $\chi_l$ & $\tilde{\chi}_l/(x_S^2/N_0 t_S)$
    & $1000$ \\
    
    $\chi_c$ & $\tilde{\chi}_c/(x_S^2/N_0 t_S)$
    & $500$ \\
    
    $\mu_a$ & $\tilde{\mu}_a/(1/t_S)$
    & $\numrange{10}{100}$ (variable) \\
    
    $\mu_e$ & $\tilde{\mu}_e/(1/N_0 t_S)$
    & $\numrange{10}{200}$ (variable) \\
    
    $\mu_p$ & $\tilde{\mu}_p/(1/N_0 t_S)$
    & $300$ \\
    
    $\sigma_f$ &  $\tilde{\sigma}_f/(x_S/t_S)$
    & $100$ \\
    
    $\sigma_l$ & $\tilde{\sigma}_l/(N_0 x_S/t_S)$
    & $10$ \\
    
    $\sigma_e$ & $\tilde{\sigma}_e/(x_S/t_S)$
    & $\numrange{10}{100}$ (variable) \\
    
    \bottomrule
    
\end{tabular}
\end{table}

The final nondimensional model consists of the constitutive equations
\begin{align}
\diffp{f}{t} &= -\diffp{}{x} (J_f + v f) + s_f \,, \label{eq:2-conteqf2} \\
\diffp{l}{t} &= -\diffp{}{x} (J_l + v l) + s_l \,, \\
\diffp{c}{t} &= -\diffp{}{x} (J_c + v c) + s_c \,,
\end{align}
with no-voids condition
\begin{equation} \label{eq:2-novoids}
f + l + c = 1 \,,
\end{equation}
source terms
\begin{alignat}{4}
s_f &= \,&+ \mu_p f l &+ \mu_e f c - \mu_a f &\,, \label{eq:2-sf} \\
s_l &= \,&- \mu_p f l & &\,, \label{eq:2-sl} \\
s_c &= &            &- \mu_e f c + \mu_a f &\,, \label{eq:2-sc} 
\end{alignat}
flux terms
\begin{align}
J_f &= - D_f \diffp{f}{x} + \chi_l f \diffp{l}{x} + \chi_c f \diffp{c}{x} \,,\\
J_l &= - D_l \diffp{l}{x} \,, \\
J_c &= - D_c \diffp{c}{x} \,,
\end{align}
boundary conditions at $x = 0$

\begin{alignat}{2}
( J_f + v f )
&= \sigma_f l \,, \\
( J_l + v l )
&= \sigma_l \,, \\
( J_c + v c ) &= 0 \,,
\end{alignat}
boundary conditions at $x = R$
\begin{alignat}{2}
( J_f + v f )
- R' f
&= \sigma_e f \,, \\
( J_l + v l )
- R' l &= 0 \,, \\
( J_c + v c )
- R' c &= 0 \,,
\end{alignat}
and initial conditions
\begin{align}
f(x,0) &= 1 \,, \\
l(x,0) &= 0 \,, \\
c(x,0) &= 0 \,, \\
R(0) &= 1 \,,
\end{align}
where the mixture velocity is
\begin{equation}
v = \sigma_f l \Big\vert_{x=0} + \sigma_l
    + D_f \diffp{f}{x} - \chi_l f \diffp{l}{x} - \chi_c f \diffp{c}{x}
    + D_l \diffp{l}{x} + D_c \diffp{c}{x} \,.
\end{equation}
and the domain grows at rate
\begin{equation}
\diff{R}{t} = \sigma_f l \Big\vert_{x=0} + \sigma_l
    - \sigma_e f \Big\vert_{x=R}\,.
\end{equation}

\subsection{Numerical solution}

The system was solved numerically using the method of lines.
The nondimensional system was first mapped onto a fixed spatial domain $[0,1]$
using the transformation $(x,t) \to (y,\tau) = (x/R(t),t)$,
thereby transforming it into a system of parabolic PDEs
coupled to an ODE for the domain length $R(\tau)$.
The transformed system was then reduced to a set of time-dependent ODEs
by discretising spatial derivatives
using a central differencing approximation.
The resulting ODEs were solved using a backward differentiation formula method
in Python using SciPy's integration libraries.

\section{Plaque structure, cell death, and efferocytosis}
\label{sec:3}

In this section we neglect emigration, setting $\sigma_e = 0$,
and focus on how the relative rates of cell death and efferocytosis
influence the composition of the plaque.

\subsection{Base case simulation}

\begin{figure}
    \centering
    \begin{tabular}{l l l}
        (a) & \includegraphics[align=t,trim={0 0 7cm 0},clip]
        {{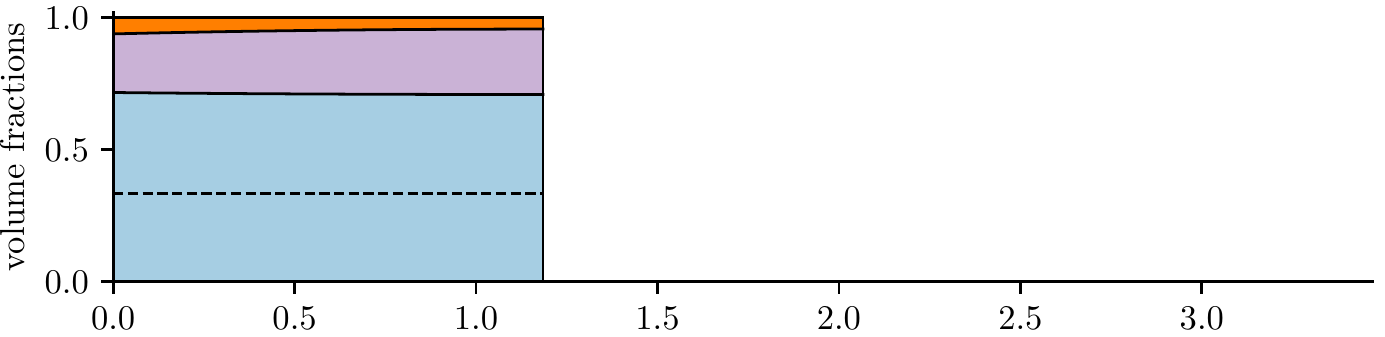}} &
        \includegraphics[align=t,height=48pt]{{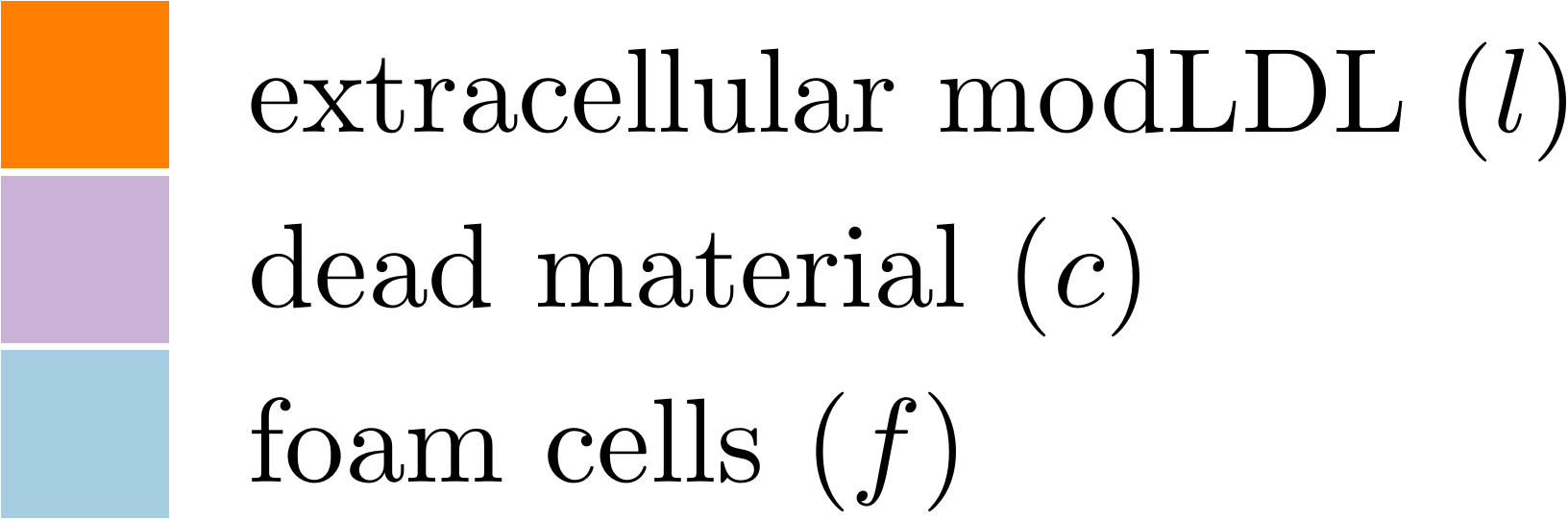}} \\
        (b) & \multicolumn{2}{l}{
            \includegraphics[align=t,trim={0 0 7cm 0},clip]
            {{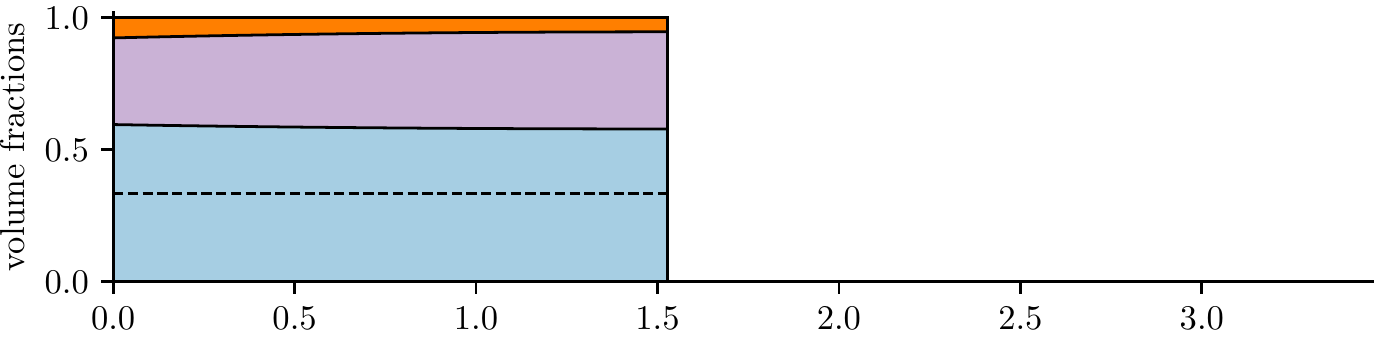}} } \\
        (c) & \multicolumn{2}{l}{
            \includegraphics[align=t]
            {{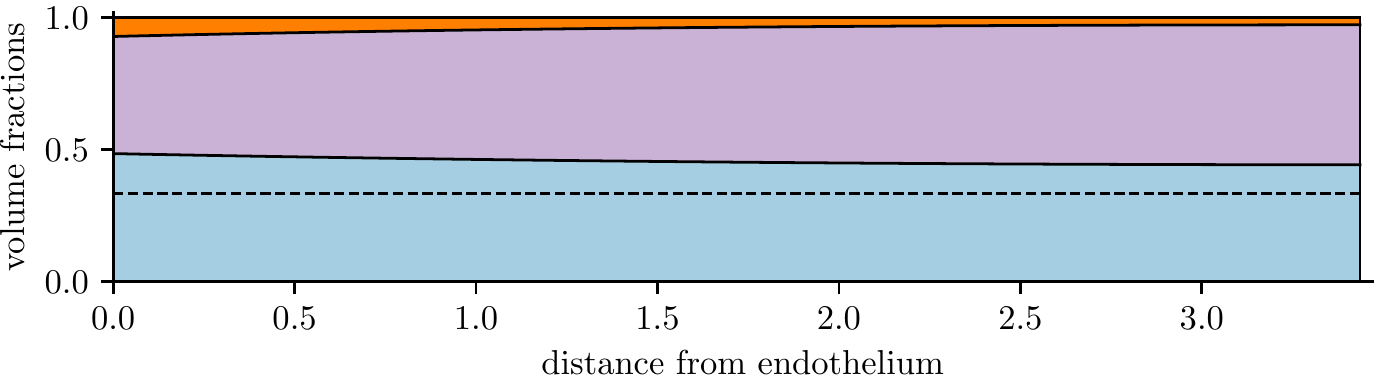}} }
    \end{tabular}
    
    \caption{Early time evolution of a plaque's structure
        with $\mu_a = 40$, $\mu_e = 60$, $\sigma_e = 0$,
        plotted at times (a) $t = 0.01$, (b) $t = 0.025$, (c) $t = 0.1$.
        The dotted line denotes the $f^* = 1 - \tfrac{\mu_a}{\mu_e} = 0.333$
        steady state foam cell density predicted by the ODE model. }
    \label{fig:3-distbase}
    \vspace{1cm}
\end{figure}

\begin{figure}
    \centering
    
    \begin{tabular}{l l l}
        (a) & \multicolumn{2}{l}{
            \includegraphics[align=t]
            {{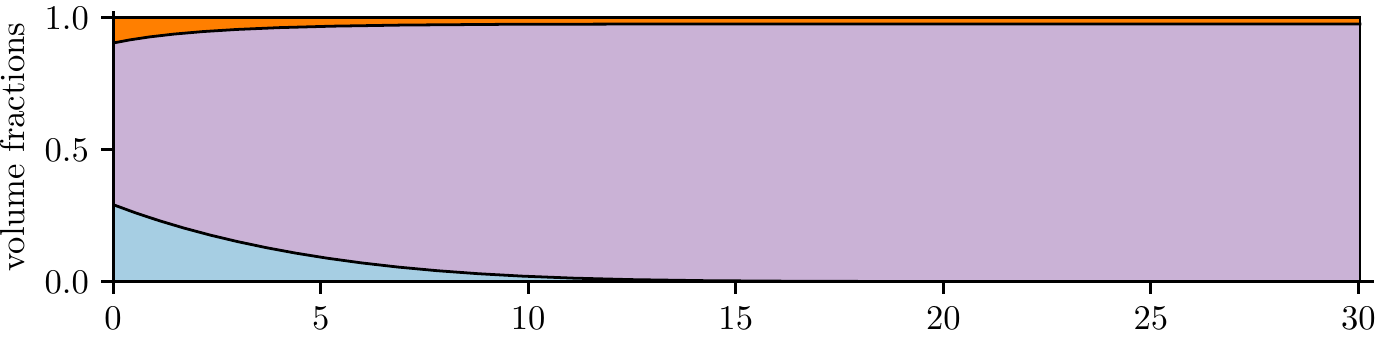}} } \\
        (b) & \multicolumn{2}{l}{
            \includegraphics[align=t]
            {{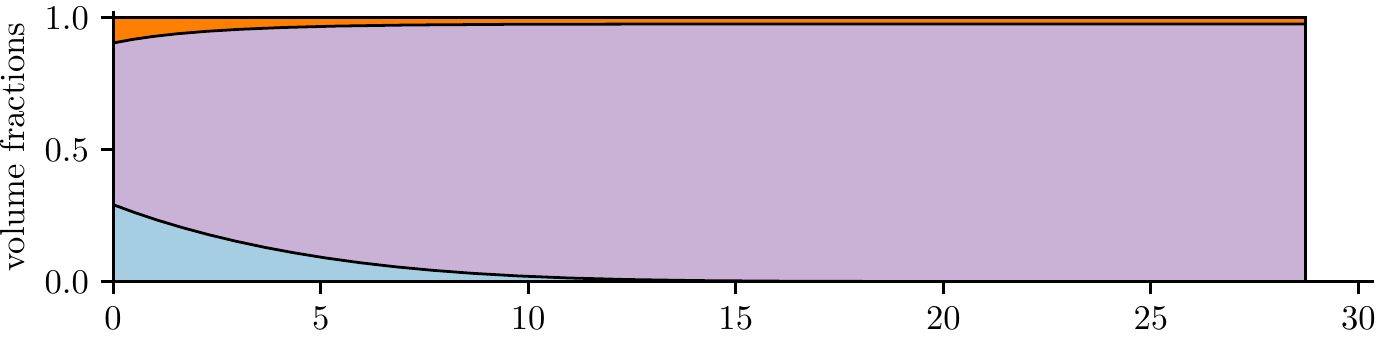}} } \\
        (c) & \multicolumn{2}{l}{
            \includegraphics[align=t]
            {{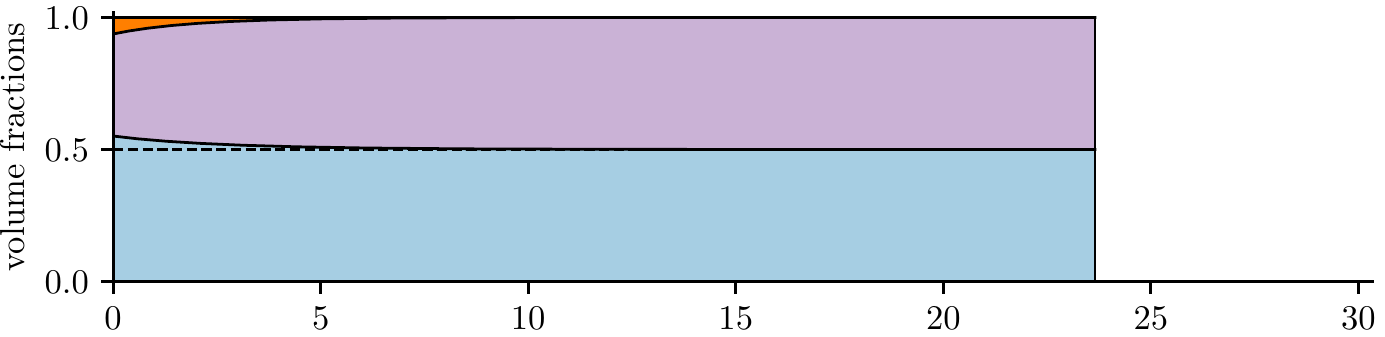}} } \\
        (d) & \includegraphics[align=t,trim={0 0 4cm 0},clip]
        {{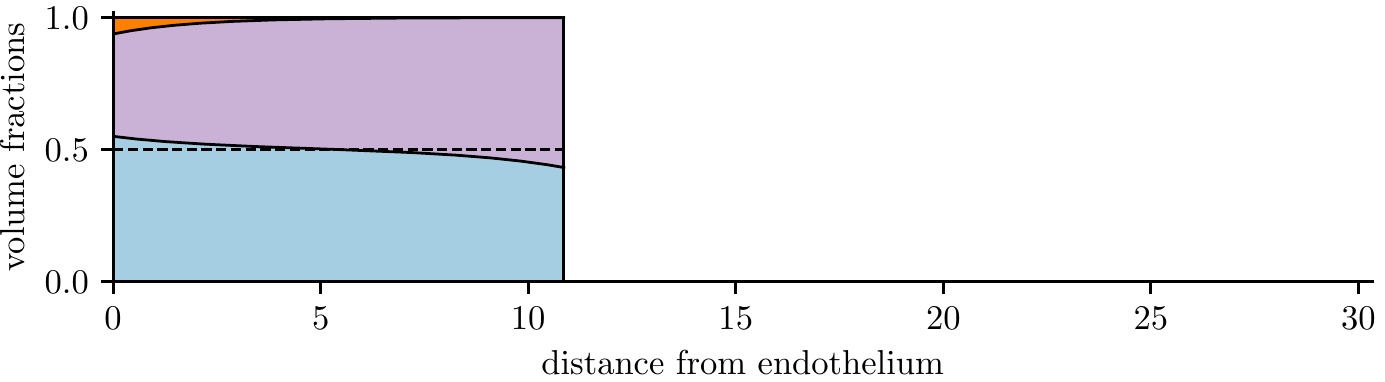}} &
        \includegraphics[align=t,height=48pt]
        {{fig_phaselegend.pdf}} \\
    \end{tabular}
    
    \vspace{2mm}
    \caption{Comparisons of plaque structure at $t = 1$ for $\mu_a = 40$ and
        (a) $\mu_e = 20$, $\sigma_e = 0$,
        (b) $\mu_e = 20$, $\sigma_e = 30$,
        (c) $\mu_e = 80$, $\sigma_e = 0$,
        (d) $\mu_e = 80$, $\sigma_e = 30$.
        The dotted lines in (c) and (d) denote the
        $f^* = 1 - \tfrac{\mu_a}{\mu_e} = 0.5$ steady state foam cell density
        predicted by the ODE model
        (for (a) and (b), an $f^* = 0$ steady state is predicted). }
    \label{fig:3-distcomp}
\end{figure}

\Cref{fig:3-distbase} illustrates the evolution of a base case with
relatively high efferocytic uptake compared to cell death.
Following the initial injury where LDL starts being deposited in the intima,
the plaque quickly reaches a state where
it is populated almost entirely by foam cells and dead material,
although the balance between the two may vary
(\Cref{fig:3-distbase}).
A small amount of modLDL that has yet to be ingested
remains near the endothelium,
though some uningested modLDL may remain throughout the intima
for very early times (\Cref{fig:3-distbase}),
or for cases where all foam cells in the deep plaque have died
(\Cref{fig:3-distcomp}(a)).
ModLDL and live foam cell densities
are higher near the endothelium than in the deep plaque
due to the constant influx of new monocytes and LDL particles,
though some dead material is still present due to diffusion.
Away from the endothelium,
phase densities are largely uniform,
with some non-uniformity observed near the medial boundary
when $\sigma_e > 0$ (\Cref{fig:3-distcomp}(d)).

\begin{figure}
\centering
    \begin{tabular}{l@{}r@{}l@{}r}
        (a) & \includegraphics[align=t]
        {{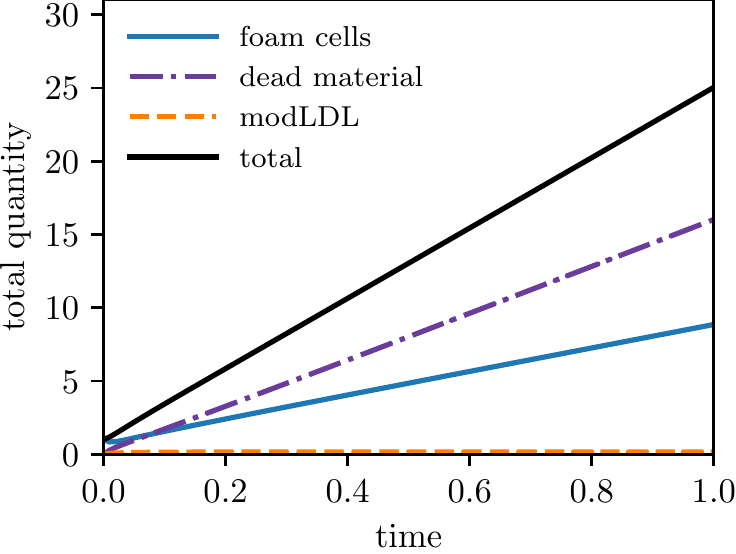}} &
        (b) & \includegraphics[align=t]
        {{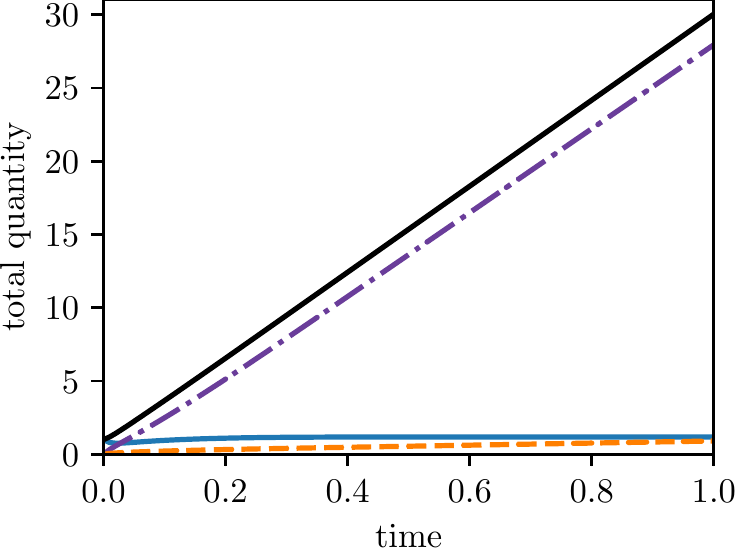}}
    \end{tabular}
    \vspace{6pt}
    \caption{Time evolution of the total phase constituents
        $\int_0^R u \dif x$ (for phases $u = f,l,c$)
        immediately following plaque initialisation.
        Plots are for plaques with $\mu_a = 40$, $\sigma_e = 0$, and
        (a) $\mu_e = 60$, (b) $\mu_e = 20$. }
    \label{fig:3-histcomp}
\end{figure}

The intima itself quickly approaches a state
where it grows at a constant rate (\Cref{fig:3-histcomp}).
The rates at which foam cells and dead material accumulate
also settle to a constant rate,
while uningested modLDL may either reach a steady value
or accumulate at a constant rate,
depending on whether the deep plaque still
has live foam cells remaining to ingest new lipid
(\Cref{fig:3-histcomp}(a) vs (b)).

We remark that
increasing the LDL deposition rate $\sigma_l$
or the monocyte recruitment parameter $\sigma_f$
will increase the intimal growth rate.
This is due to the increased influx of new material
in the form of modLDL or recruited monocytes respectively.
As we see later however,
the qualitative spatial structure
depends primarily on the cell death and efferocytosis rates,
which remain our focus in this section.

\begin{figure}
    \centering
    \begin{tabular}{l c}
        (a) & \includegraphics[align=t]
        {{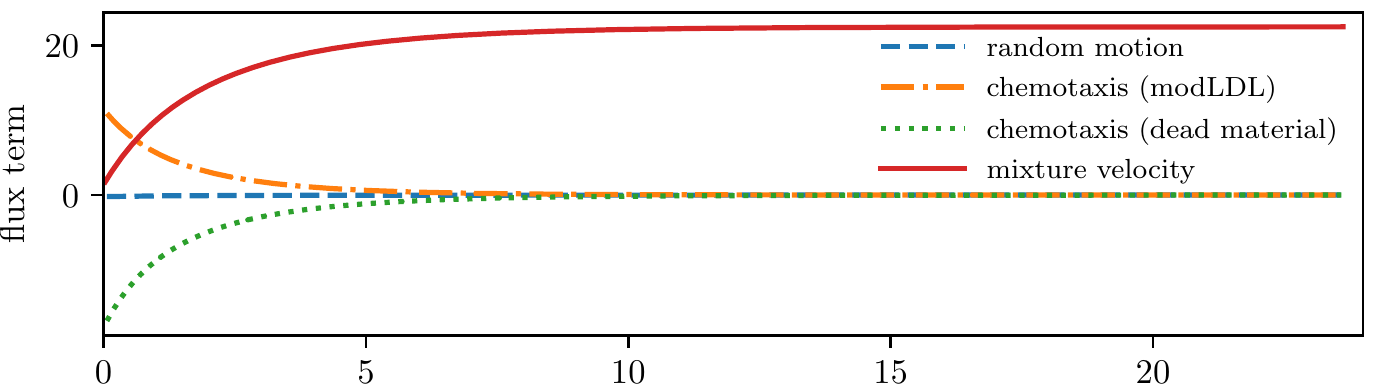}} \\
        (b) & \includegraphics[align=t]
        {{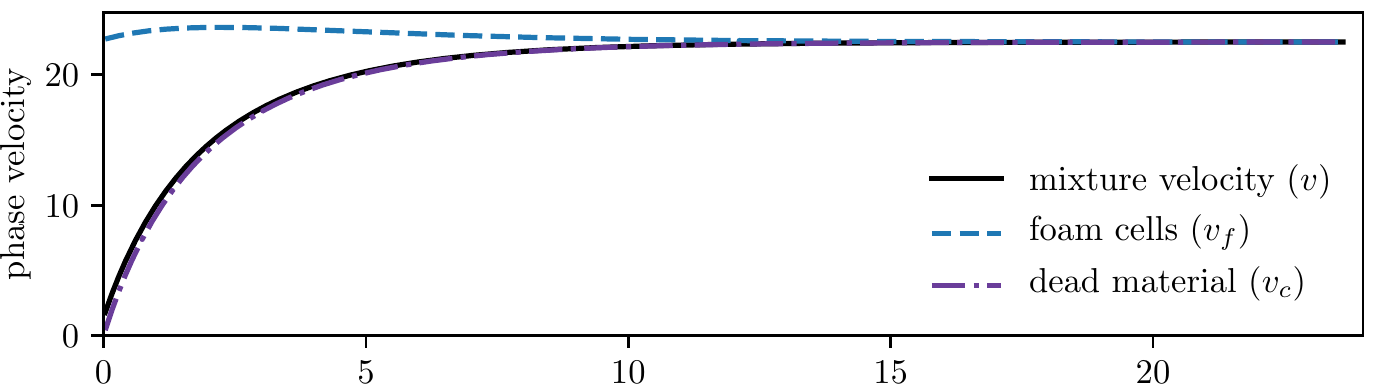}} \\
        (c) & \includegraphics[align=t]
        {{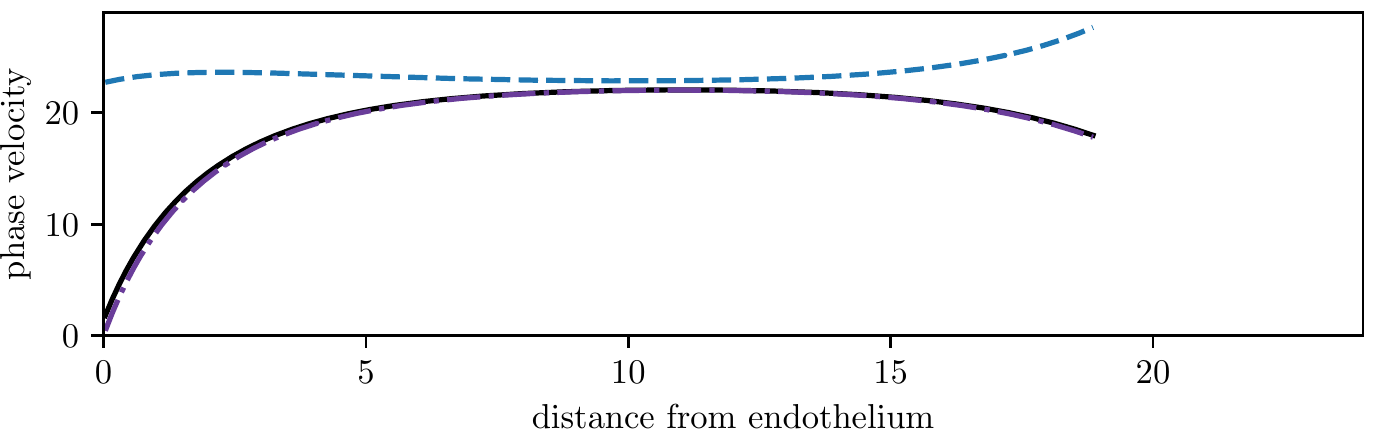}} \vspace{3mm}
    \end{tabular}
    \caption{ (a) Comparison of the individual flux terms
        in the constitutive equation (\cref{eq:2-conteqf2,eq:2-sf}))
        for the foam cell phase:
        random motion $-D_f \frac{\partial f}{\partial x}$,
        chemotaxis towards modLDL $\chi_l f \frac{\partial l}{\partial x}$
        and dead material $\chi_c f \frac{\partial c}{\partial x}$,
        and bulk advection $v f$.
        (b,c) Comparison of the foam cell and dead material phase velocities
        with the mixture velocity.
        Terms are plotted as a function of $x$ at a fixed time $t = 1$
        for a scenario with $\sigma_a = 40$
        and (b) $\sigma_e = 0$, (c) $\sigma_e = 80$.}
    \label{fig:3-flux}
\end{figure}

\Cref{fig:3-flux}(a) compares the relative sizes of
the diffusive, chemotactic, and advective foam cell flux terms.
Diffusion and chemotaxis are dominated
by advective transport, which is the sole driver
of foam cell movement in the deep plaque.
Chemotaxis is important near the endothelium
where there are larger amounts of uningested modLDL,
but becomes irrelevant in the deeper plaque
once the available modLDL has been internalised.
From \Cref{fig:3-flux}(b),
the total phase velocities for both the foam cell and dead cell phases
converge to the mixture velocity in the deep plaque,
with little interphase motion.
This suggests that mass transport in the deep plaque
occurs primarily via bulk advection.
We note however that when $\sigma_e > 0$,
emigration of foam cells into the lymphatics
causes some deviation between the mixture velocity
and the foam cell phase velocity
(\Cref{fig:3-flux}(c)).
Nevertheless, the $\sigma_e = 0$ case provides
a good baseline for comparison,
since the close agreement between the phase velocities
makes the system amenable to analysis using
the method of characteristics.

\subsection{ODE approximation for deep plaque composition}

The constitutive equations for each phase $u$ can be written in the form
\begin{equation}
\diffp{u}{t} + \diffp{}{x} (v_u u) = s_u \,,
\end{equation}
where $v_u = (J_u + v u)/u$ is the total velocity of phase $u$,
and $s_u$ is its source term.
In \Cref{fig:3-flux}, we observed that away from the endothelium,
the bulk advection term $v u$ dominates over
random and chemotactic motion modelled by $J_u$
for all three phases,
and so $v_u \approx v$ in the deep plaque.
For each phase $u$, we define $U_i(t)$ to be the phase density
along a characteristic curve $x_c(t)$ advected with speed $v_u$,
i.e.
\begin{equation}
U(t) = u( x_c(t), t ) \,,\quad \diff{x_c}{t} = v_u \,.
\end{equation}
In the deep plaque, $v_u \approx v$ for all phases,
and we can make the approximation
\begin{equation} \label{eq:3-chardUidt}
\diff{U}{t} \approx s_u \,.
\end{equation}
We note also that in the deep plaque,
the modLDL density $l(x,t)$ is negligible,
so we make an additional approximation $l = 0$.
Substituting the model source terms \labelcref{eq:2-sf,eq:2-sl}
into \cref{eq:3-chardUidt} yields the ODE system
\begin{alignat}{4}
\diff{F}{t} &= \,&- \mu_a F + \mu_e F C &\,, \label{eq:3-odedfdt} \\
\diff{C}{t} &= \,&+ \mu_a F - \mu_e F C &\,, \label{eq:3-odedcdt}
\end{alignat}
where $F(t) = f(x_c(t),t)$ and $C(t) = c(x_c(t),t)$
denote the phase densities along characteristic curves.
Due to the no-voids condition \labelcref{eq:2-novoids},
we also have $F + C = 1$.
Using this condition, the coupled equations reduce to a single ODE
\begin{equation}
\diff{F}{t} = \mu_e F \del{ \del[1]{1-\tfrac{\mu_a}{\mu_e}} - F } \,.
\end{equation}
This equation has two steady states at
$F^* = 0$ and $F^* = 1-\tfrac{\mu_a}{\mu_e}$,
with a transcritical bifurcation occurring when $\mu_a = \mu_e$.
Linearising about the steady states gives eigenvalues of
$\lambda = \mu_e - \mu_a$ at $F^* = 0$ and
$\lambda = \mu_a - \mu_e$ at $F^* = 1 - \frac{\mu_a}{\mu_e}$.

When $\mu_a < \mu_e$ and efferocytosis rates are high relative to cell death,
$F^* = 1-\tfrac{\mu_a}{\mu_e}$ is the sole stable steady state,
and the system tends towards a state where
live foam cells coexist with dead material.
When $\mu_a < \mu_e$ however, $F^* = 0$ is the sole stable state.
Here, efferocytic recycling is insufficient to counterbalance cell death,
and the plaque settles to a state where all foam cells die off.
From the eigenvalues,
for a fixed ratio $\frac{\mu_a}{\mu_e}$,
higher overall rates of cell death and efferocytosis
will cause the system to settle to the relevant steady state faster.

\subsection{Plaque composition and the death/efferocytosis balance}

\begin{figure}
    \centering
    \includegraphics[align=t]
    {{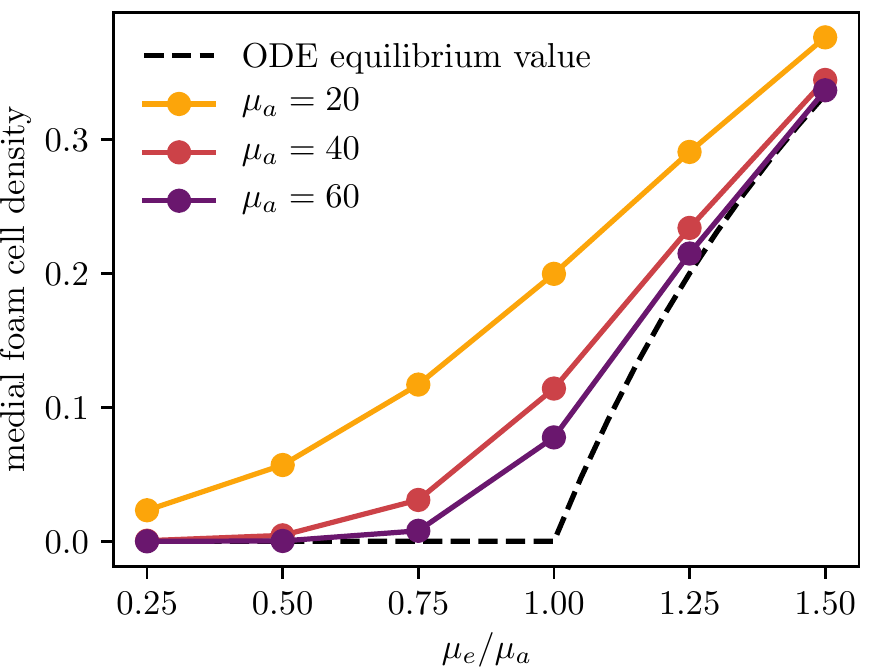}}
    \caption{Comparison of the deep plaque foam cell density
        $f|_{x=R}$ as a function of
        the efferocytosis/death ratio $\tfrac{\mu_e}{\mu_a}$,
        plotted at $t = 0.5$ for various values of $\mu_a$.
        The equilibrium density predicted by the ODE model
        is also given for comparison. }
    \label{fig:3-fR}
\end{figure}

The deep plaque composition closely matches
the steady state densities predicted by the ODE model.
In \Cref{fig:3-distcomp},
lower rates of efferocytosis are associated with
higher amounts of dead material and fewer living foam cells.
For plaques with efficient efferocytosis
where $\mu_e > \mu_a$
(\Cref{fig:3-distcomp}(c) and (d)),
the foam cell density settles to
the $f^* = 1 - \frac{\mu_a}{\mu_e}$ density
predicted by the ODE model.
On the other side of the $\tfrac{\mu_e}{\mu_a} = 1$ bifurcation point,
plaques with $\mu_e < \mu_a$
(\Cref{fig:3-distcomp}(a) and (b))
reach a state where the deep plaque supports no live foam cells.
For especially low values of $\mu_e$
(\Cref{fig:3-distcomp}(a)),
foam cells will die before ingesting all available extracellular modLDL,
leaving some remnant modLDL
which gets carried advectively into the deep plaque.
From \Cref{fig:3-fR},
higher rates of death and efferocytosis
for a given value of $\frac{\mu_a}{\mu_e}$
lead to better agreement between the ODE steady states
and the deep plaque densities at $x = R$.
This is consistent with linear analysis of the ODE system,
which predicts faster convergence to the steady state
for higher values of $\mu_a$ and $\mu_e$.

\begin{figure}
    \centering
    (a) \includegraphics[align=t]
    {{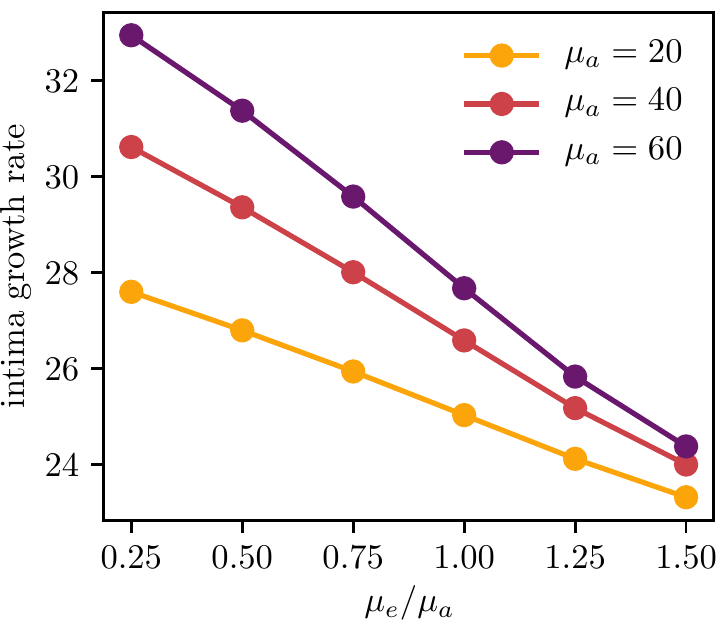}}
    (b) \includegraphics[align=t]
    {{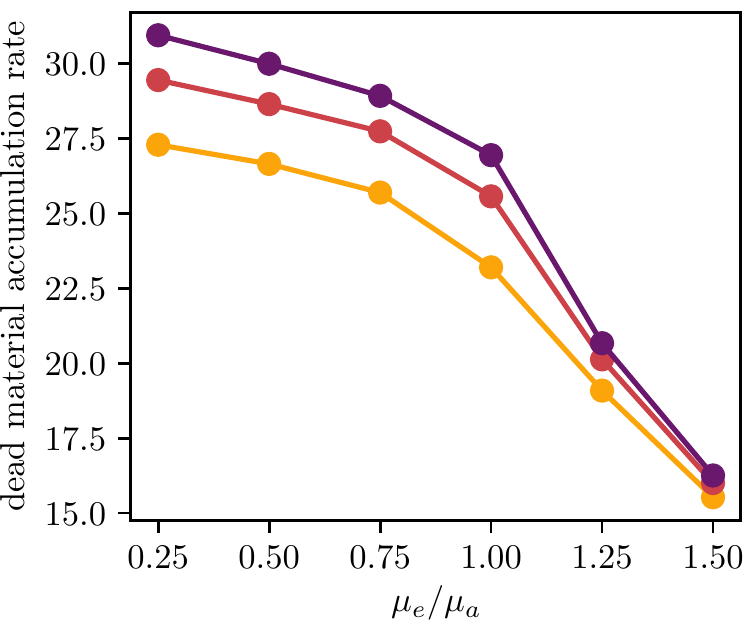}}
    \caption{Comparison of
        (a) the intima growth rate $\frac{\dif R}{\dif t}$ and
        (b) the rate of change of total dead material
        $\frac{\dif}{\dif t} \int_0^R c \,\dif x$
        as a function of the efferocytosis/death ratio $\tfrac{\mu_e}{\mu_a}$,
        plotted at $t = 1$ for various values of $\mu_a$. }
    \label{fig:3-growthaccum}
\end{figure}

Cell death and efferocytosis also influence
the growth rate of a plaque,
even with fixed values of the monocyte and LDL influx parameters
$\sigma_f$ and $\sigma_l$.
From \Cref{fig:3-growthaccum}(a),
increasing $\mu_e$ with fixed $\mu_a$
will slow the rate of growth.
This is because increasing the efficiency of efferocytosis
will increase the numbers of foam cells
actively consuming modLDL,
thus reducing endothelial modLDL levels
and dulling the inflammatory response.
This slowing in growth
can also be observed when comparing plaque sizes
in \Cref{fig:3-distcomp}.
Reducing $\mu_e$ and $\mu_a$ with constant $\frac{\mu_e}{\mu_a}$
has the same effect of slowing plaque growth
by reducing endothelial modLDL levels,
since reducing the rate at which cells start to die
will increase the quantity of available foam cells
near the endothelium.

The rate at which a plaque accumulates new dead material
is a useful marker of plaque health
alongside its total growth rate.
Both are low in a healthy plaque.
As discussed previously,
reducing $\mu_e$ and $\mu_a$ with constant $\frac{\mu_e}{\mu_a}$
reduces the rate at which new cellular material enters the plaque.
From \Cref{fig:3-growthaccum}(b),
this results in a slight decrease
in the total rate of increase of dead material,
although the proportion of dead material
in the bulk of the plaque remains the same.
Increasing the efficiency of efferocytosis via $\mu_e$
with constant $\mu_a$
has a more significant effect on the rate of accumulation,
as it reduces both the monocyte influx
and the local dead cell proportion
in the interior of the plaque
by promoting more active uptake of dead material.

\section{Emigration and plaque growth}
\label{sec:4}

Foam cell emigration can slow plaque growth
and potentially allow plaque size to stabilise
by allowing material to leave the plaque.
The ability of emigration to impact plaque growth however
depends on the availability of live foam cells.

\begin{figure}
    \centering
    (a)\includegraphics[align=t]
    {{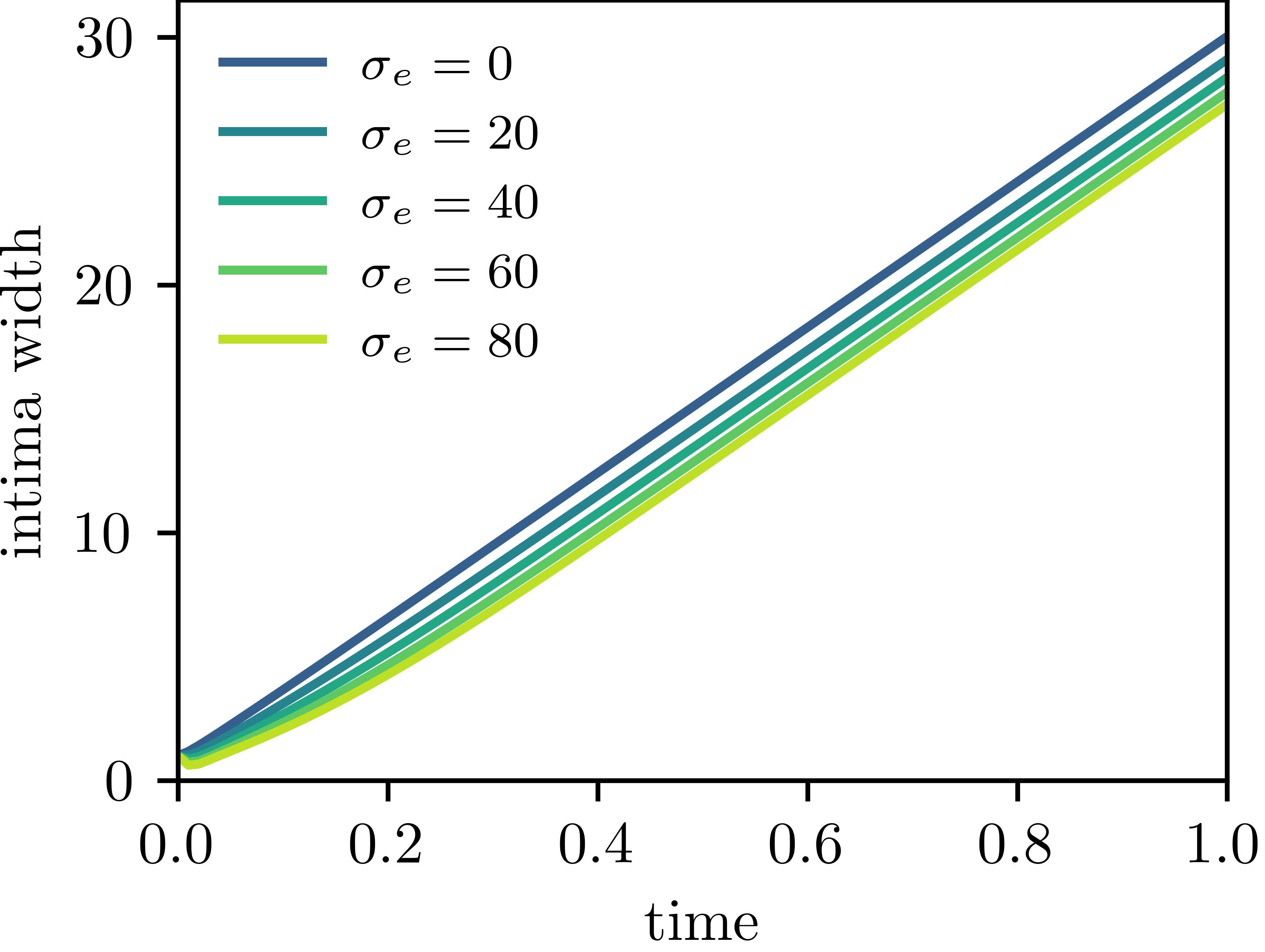}}
    (b) \includegraphics[align=t]
    {{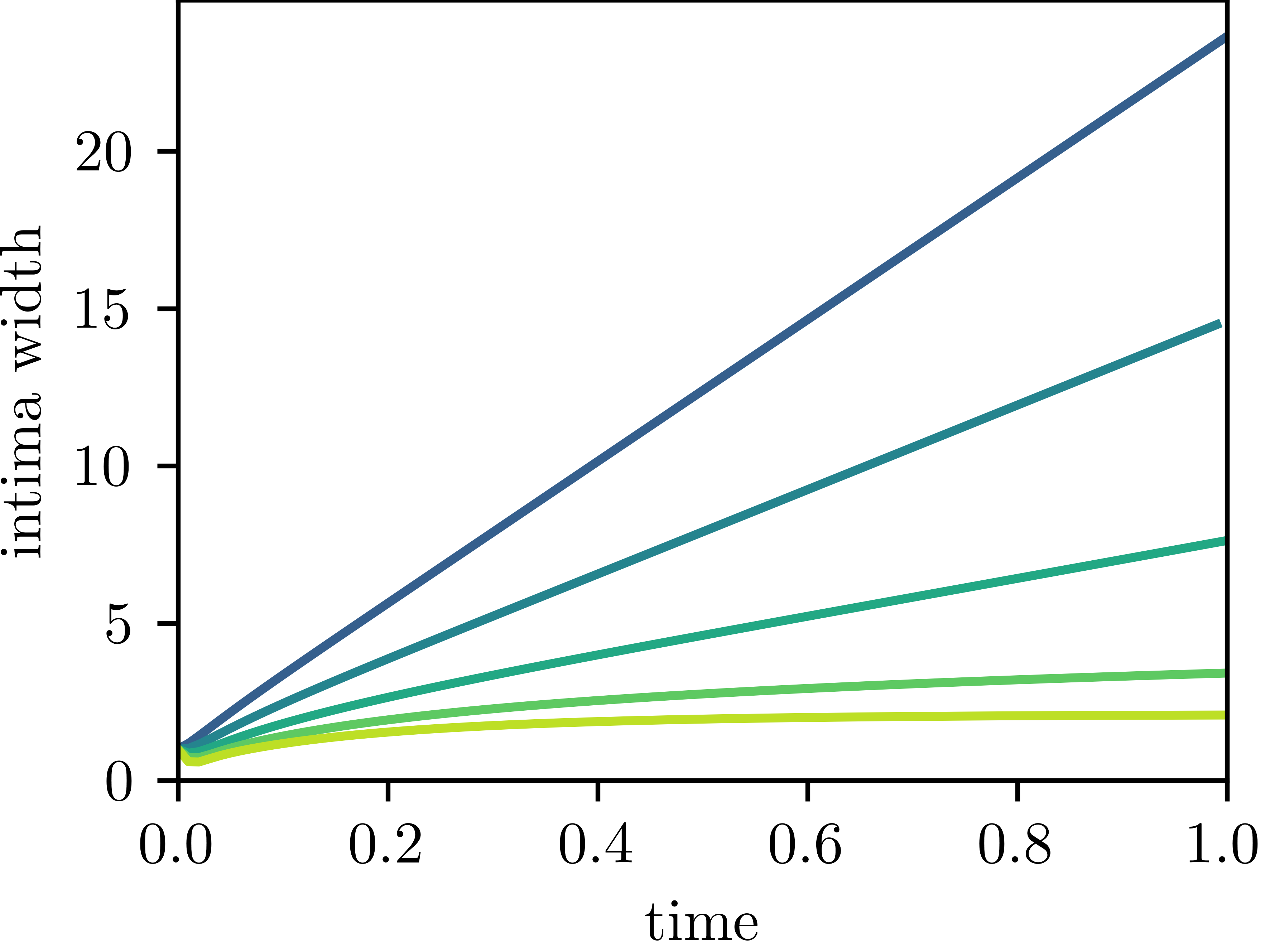}}
    \vspace{6pt}
    \caption{Plaque growth for varying egress velocities $\sigma_e$,
        for low and high $\tfrac{\mu_e}{\mu_a}$ scenarios.
        Plots track the intima width $R(t)$, where $\mu_a = 40$, and
        (a) $\mu_e = 20$, (b) $\mu_e = 80$. }
    \label{fig:4-Regress}
\end{figure}

In \Cref{fig:3-distcomp}(a,b) and \labelcref{fig:4-Regress}(a),
$\mu_e < \mu_a$,
and efferocytosis is insufficient to prevent
the death of foam cells in the deep plaque.
Increasing the emigration velocity here
has little effect on plaque size
as there are few foam cells close to the medial edge of the plaque.
In \Cref{fig:3-distcomp}(c,d) on the other hand, $\mu_e > \mu_a$,
and there is a higher density of foam cells in the deep plaque.
Increasing the emigration velocity here
slows the rate of plaque growth
by enabling material to leave the plaque,
which also results a reduction in plaque size.
For sufficiently high $\sigma_e$,
the plaque will stabilise entirely and settle to
a steady state with no growth (\Cref{fig:4-Regress}(b)).
While emigration does cause a slight reduction in
foam cells numbers near the media,
their numbers remain sufficient to allow emigration.
We remark briefly that emigration causes
phase densities near $x = R$
to deviate from those predicted by the ODE model,
which we discuss further in \Cref{sec:discussion}.

\begin{figure}
    \centering
    \hspace{10mm} (a) \includegraphics[align=t]
    {{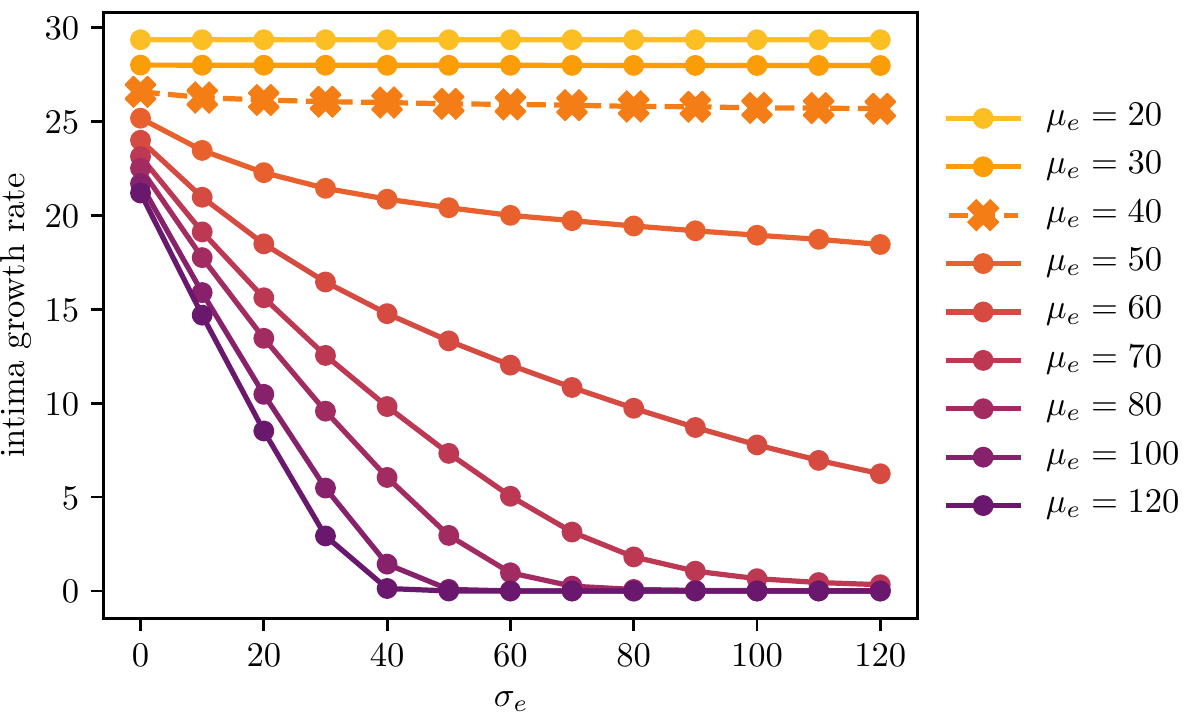}} \\ \vspace{2mm}
    (b) \includegraphics[align=t]
    {{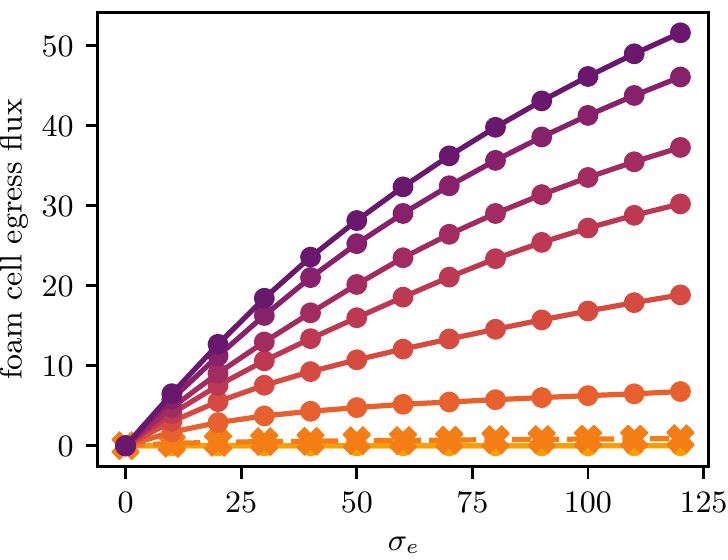}}
    (c) \includegraphics[align=t]
    {{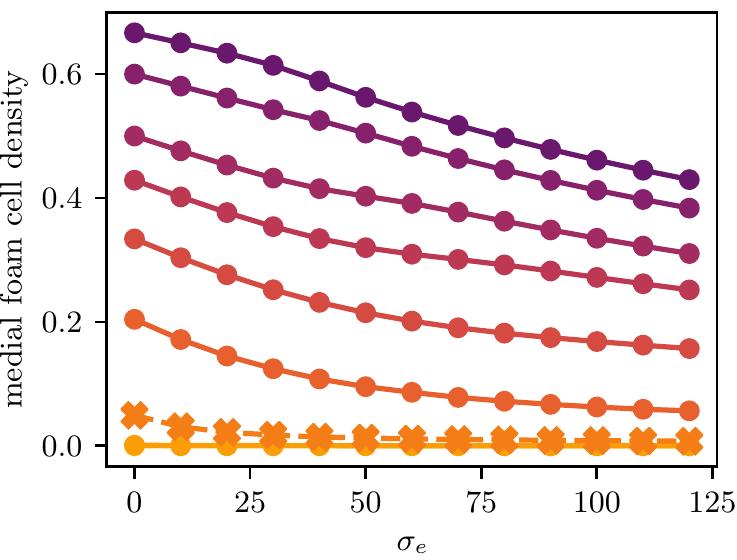}}
    \vspace{2mm}
    \caption{ Plots showing the effect of the emigration velocity $\sigma_e$
        on plaque development,
        compared across various efferocytosis rates $\mu_e$
        with $\mu_a = 40$ fixed.
        Quantities are measured at $t = 1$.
        Plots compare
        (a) the total intima growth rate $\frac{\dif R}{\dif t}$,
        (b) the total flux of foam cells out of the medial boundary
        $j_f|_{x=R}$ via emigration, and
        (c) the foam cell density at the medial boundary $f|_{x=R}$.
    }
    \label{fig:4-egressqties}
\end{figure}

The importance of healthy efferocytosis
in enabling foam cell emigration is more apparent
when considering a range of values of $\mu_e$.
From \Cref{fig:4-egressqties}(b),
increasing the emigration velocity $\sigma_e$
will increase the total foam cell emigration flux
for most values of $\mu_e$.
This flux is diminished for lower values of $\mu_e$,
and for $\mu_e \leq \mu_a$,
foam cell emigration remains negligible
even for very high values of $\sigma_e$.
\Cref{fig:4-egressqties}(c) suggests that
for cases where when $\mu_e \leq \mu_a$,
the lack of emigration observed is due to
the absence of live foam cells in the deep plaque.
Conversely, cases with $\mu_e > \mu_a$ will still have
some foam cells near the media,
thus enabling emigration.
Even for these cases however,
higher emigration velocities result in lower medial foam cell densities
due to the faster removal of foam cells.
In particular, for cases where $\mu_e > \mu_a$
with $\mu_e$ close to $\mu_a$,
the low live foam cell density
gives diminishing returns on the total emigration flux
when the emigration velocity is increased.

These results suggest that
cell death will interfere with or prevent
plaque stabilisation via foam cell emigration,
depending on how efficiently efferocytosis acts.
From \Cref{fig:4-egressqties}(a),
for $\mu_e \leq \mu_a$,
increasing $\sigma_e$ has no effect
on the plaque growth rate
as there are no foam cells to emigrate from the plaque.
For values of $\mu_e > \mu_a$ close to $\mu_a$,
emigration is able to slow plaque growth,
but full plaque stabilisation is not observed
even for very high values of $\sigma_e$.
For higher values of $\mu_e$,
stabilisation is observed for high enough $\sigma_e$,
with higher $\mu_e$ scenarios
achieving this more readily.
We briefly note that the endothelial LDL and monocyte influx parameters
$\sigma_l$ and $\sigma_f$ also affect plaque growth
in the presence of emigration.
This is because higher $\sigma_l$ and $\sigma_f$
necessitate higher emigration rates $\sigma_e$
to balance the higher rates of material influx at the endothelium.

\section{Macrophage circulation and retention}
\label{sec:5}

In this section, we investigate the movement and retention
of macrophage foam cells in the intima.
We do so by extending the model to include an additional bead species
that moves with foam cells (or dead cells),
based on experimental work that labels and tracks macrophages
using fluorescent latex microspheres \citep{williams2018limited}.

\subsection{Extending the model: bead tagging}

For the bead labelling model,
we add two new species $q_f$ and $q_c$,
representing beads carried by live foam cells and dead material respectively.
We assume that the volume occupied by the beads is negligible,
and that their densities can be modelled by the continuity equations
\begin{align}
\diffp{q_f}{t} &= -\diffp{}{x} (J_{q_f} + v q_f) + s_{q_f} \,, \\
\diffp{q_c}{t} &= -\diffp{}{x} (J_{q_c} + v q_c) + s_{q_c} \,.
\end{align}
We suppose that beads are transported with the same velocity as
the phase in which they are being carried.
Using the flux-velocity relation $v_u u = j_u \,(= J_u + v u)$
gives the bead flux terms
\begin{alignat}{3}
J_{q_f} &= \frac{q_f}{m} J_f \,,\quad&
J_{q_c} &= \frac{q_c}{c} J_c \,. \label{eq:5-Jbeads}
\end{alignat}
Beads carried by macrophages are released into the dead material phase
at the rate $\mu_a$ with which foam cells die.
Similarly, beads within dead material
are taken up by foam cells via efferocytosis
at the same rate $\mu_e m$ as dead material is taken up.
This gives the bead source terms
\begin{align}
s_{q_f} &= -\, \mu_a q_f + \mu_e m q_c \,, \\
s_{q_c} &= +\, \mu_a q_f - \mu_e m q_c \,.
\end{align}
Note that bead numbers are locally conserved since $s_{q_f} + s_{q_c} = 0$.

At the endothelial boundary, we assume that
macrophage-carried beads are introduced over
a brief time period centred at $t_{\text{tag}}$.
We model the bead influx with a Gaussian function,
so that bead counts rise and fall over
approximately $6$ hours ($t = 0.035$).
We represent this via the following bead endothelial boundary conditions:
\begin{align}
(J_{q_f} + v q_f) \Big\vert_{x=0} &=
    \exp\del[2]{ -\tfrac{1}{2} (\tfrac{t-t_{\text{tag}}}{0.035})^2 } \,, \\
(J_{q_c} + v q_c) \Big\vert_{x=0} &= 0 \,.
\end{align}
The corresponding medial boundary conditions are
\begin{align}
(J_{q_f} + v q_f) \Big\vert_{x=R} - R' q_f &=
    \sigma_e q_f \Big\vert_{x=R} \,, \\
(J_{q_c} + v q_c) \Big\vert_{x=R} - R' q_c &= 0 \,.
\end{align}
These equations can be derived from \cref{eq:5-Jbeads,eq:2-bcRc}.

\subsection{Results}

\begin{figure}
    \centering
    \begin{tabular}{l l}
        (a) & \includegraphics[align=t]
            {{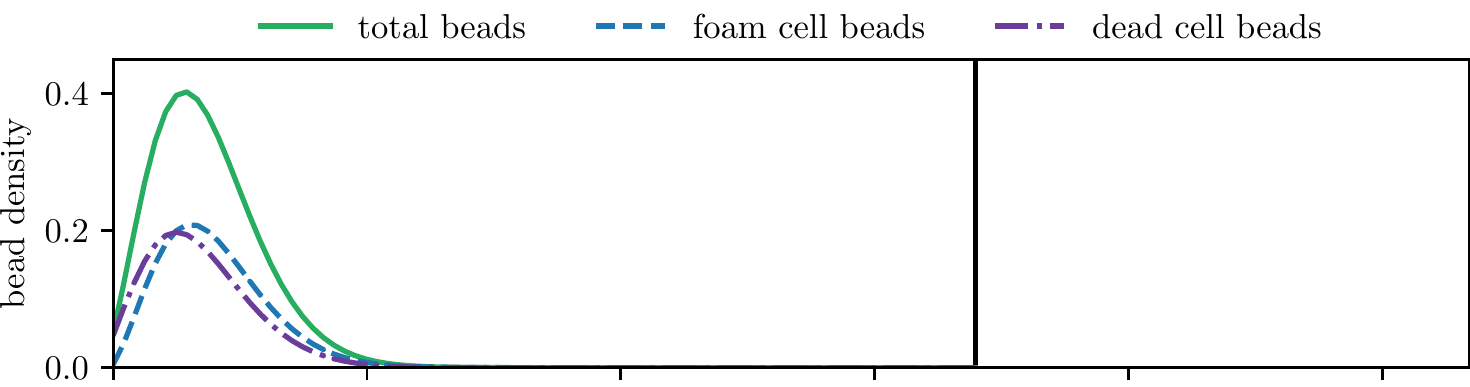}} \\
        (b) & \includegraphics[align=t]
            {{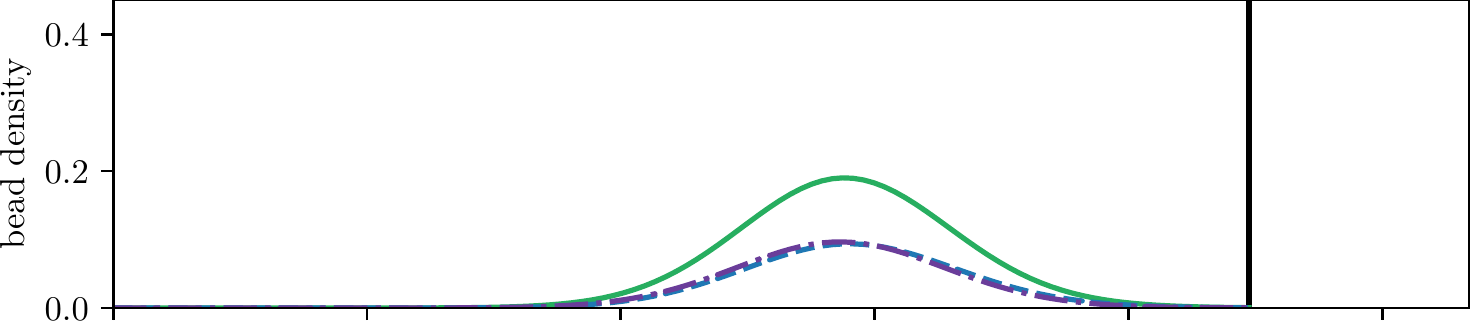}} \\
        (c) & \includegraphics[align=t]
            {{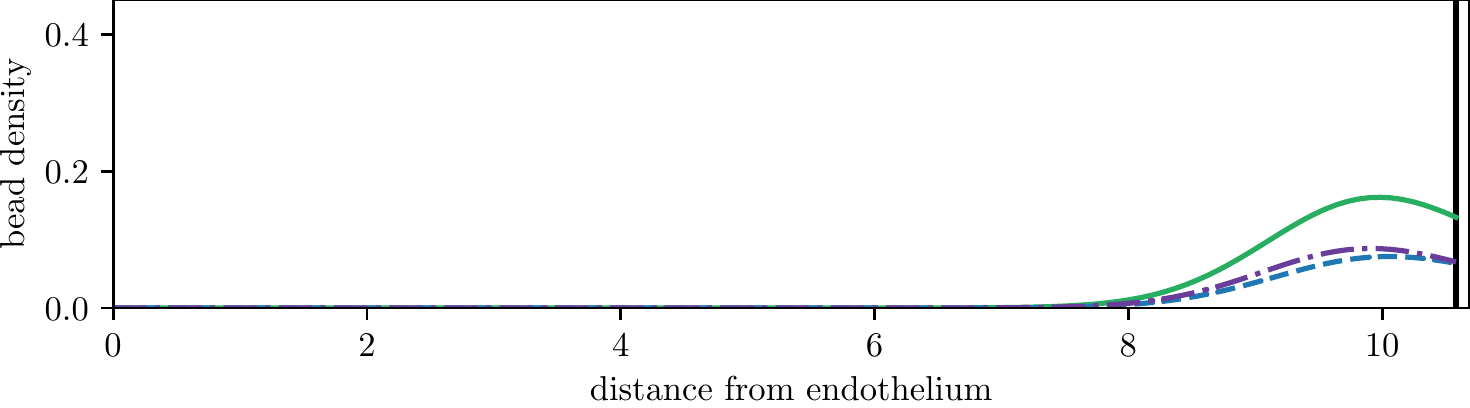}}
    \end{tabular}
    
    \caption{Time evolution of the bead spatial distribution
        for a plaque with high efferocytosis
        ($\mu_a =40$, $\mu_e = 80$, $\mu_e =20$),
        where the bead tag dosage is centred at time $t_{\text{tag}} = 1.0$.
        Distributions are plotted at times
        (a) $t = 1.1$, (b) $t = 1.5$, and (c) $t = 1.8$.
        The solid black line denotes the medial plaque boundary $x=R(t)$. }
    \label{fig:5-beadevol}
\end{figure}

\begin{figure}
    \centering
    (a) \includegraphics[align=t]
    {{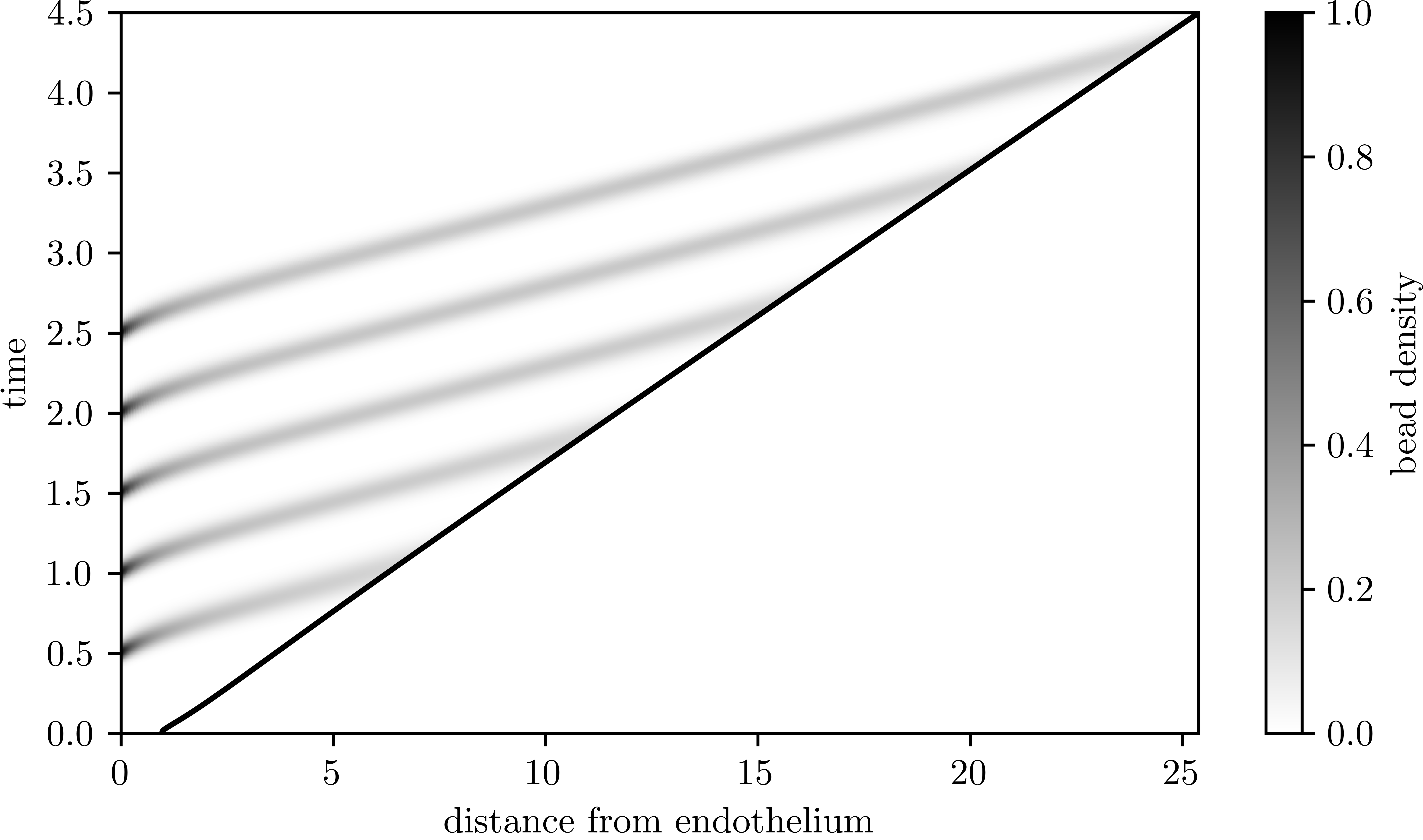}} \\ \vspace{2mm}
    (b) \includegraphics[align=t]
    {{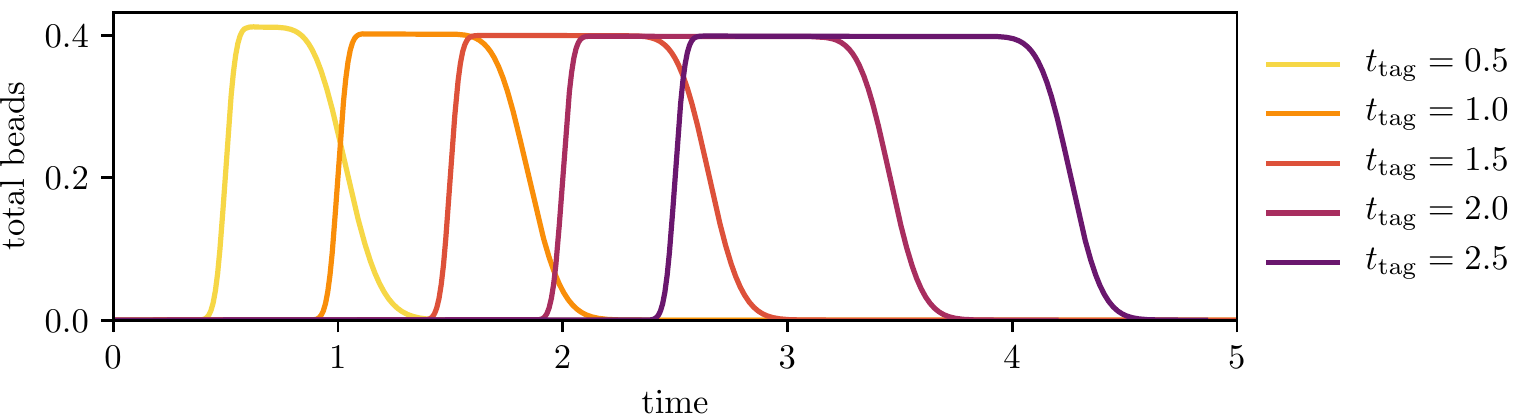}}
    \vspace{2mm}
    \caption{ Bead circulation for different bead tagging times
        for a plaque with high efferocytosis
        ($\mu_a = 40$, $\mu_e = 80$, $\sigma_e = 20$).
        Plot (a) shows the time evolution of the total bead distribution
        $q_f(x,t) + q_c(x,t)$,
        where beads are administered multiple times at
        $t_{\text{tag}} = 0.5,\, 1.0,\, 1.5,\, 2.0,$ and $2.5$.
        Plot (b) compares the time evolution of
        the total amount of beads $Q(t) = \int_0^{R(t)} (q_f+q_c) \dif x$
        for each separate bead dosage.
    }
    \label{fig:5-pathseffhigh}
\end{figure}

For plaques with sufficiently high foam cell emigration,
foam cells will move closer to the medial boundary
and eventually reach it,
before emigrating into the arterial lymphatics.
\Cref{fig:5-beadevol} and \Cref{fig:5-pathseffhigh}(a)
show the time evolution of the bead distribution
in a plaque with nonzero emigration ($\sigma_e = 20$)
and efferocytosis rates that allow
live foam cells to persist deep into the plaque
($\mu_a = 40$, $\mu_e = 80$, so $\frac{\mu_e}{\mu_a}=2$).
After bead-carrying monocytes have been recruited,
the bead population is pushed deeper into the plaque as the plaque grows.
Despite the increasing plaque size,
the distance from the bead population to the medial boundary
decreases due to material exiting through the boundary.

\begin{figure}
    \centering
    \includegraphics[align=t]
    {{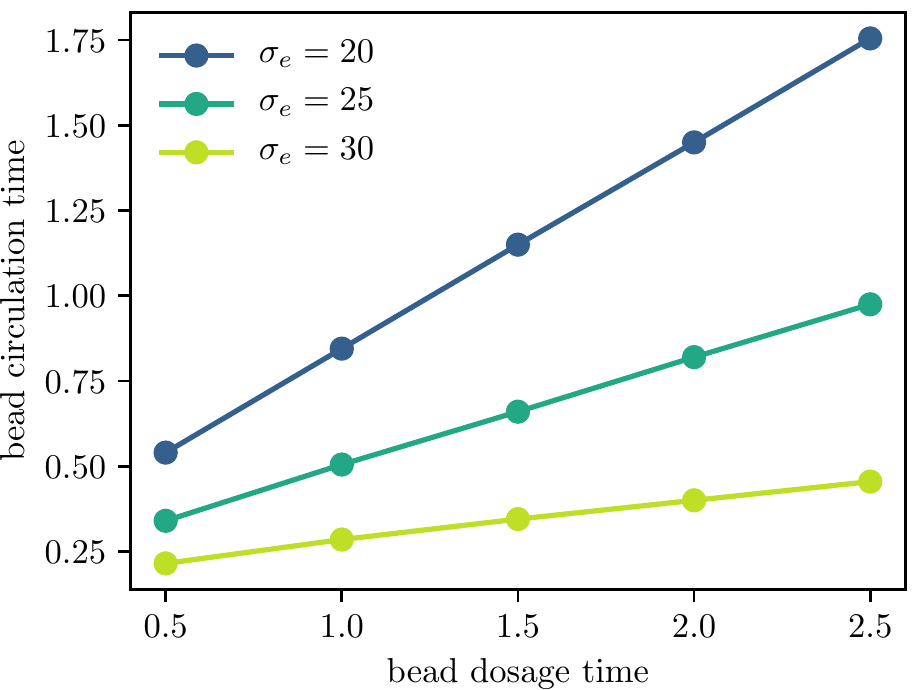}}
    \vspace{2mm}
    \caption{ Total bead circulation time $t_{\text{circ}}$
        for varying bead tagging times $t_{\text{tag}}$,
        plotted for different emigration velocities $\sigma_e$.
        Plaque parameters used are $\mu_a = 40$, $\mu_e = 80$.
    }
    \label{fig:5-circtimes}
\end{figure}

The amount of time that foam cells spend in the intima
depends on emigration rates and the size of the plaque.
\Cref{fig:5-pathseffhigh} considers the same
high efferocytosis plaque scenario as \Cref{fig:5-beadevol},
but with bead tagging happening at different times.
From \Cref{fig:5-pathseffhigh}(a),
as the plaque increases in size,
tagged foam cells that are recruited later
have to travel a greater distance before they reach the medial boundary.
As a result, they spend more time in the plaque.
\Cref{fig:5-pathseffhigh}(b) shows
the time evolution of the total amount of beads
\begin{equation}
Q(t) = \int_0^{R(t)} (q_f(x,t) + q_c(x,t)) \dif x \,.
\end{equation}
Bead totals peak soon after the tagging time $t_{\text{tag}}$,
and fall to 0 when tagged foam cells emigrate from the intima.
Bead-tagged foam cells that are recruited later
spend more time in the intima.
\Cref{fig:5-circtimes} shows how
the total circulation time of tagged monocytes depends on
the time of recruitment and the average emigration velocity.
The total circulation time $t_{\text{circ}}$ is estimated using the
temporal full width at half maximum duration of the peak bead quantity, i.e.
\begin{equation} \label{eq:5-circtime}
t_{\text{circ}} =
\max\cbr[1]{ t \,|\, Q(t) \geq 0.5 \, \max_t Q(t) }
- \min\cbr[1]{ t \,|\, Q(t) \geq 0.5 \, \max_t Q(t) } \,.
\end{equation}
Tagged monocytes that are recruited later circulate for longer time periods.
Increased migratory behaviour of foam cells
leads to reduced circulation times.
This is because higher values of $\sigma_e$ slow plaque growth,
reducing the plaque size and
the distance foam cells have to travel before exiting the plaque.
As observed earlier in \Cref{fig:3-flux}(b) and (c),
higher values of $\sigma_e$ also increase the velocity
of foam cells near the medial boundary.

\begin{figure}
    \centering
    (a) \hspace{0.7cm} \includegraphics[align=t]
    {{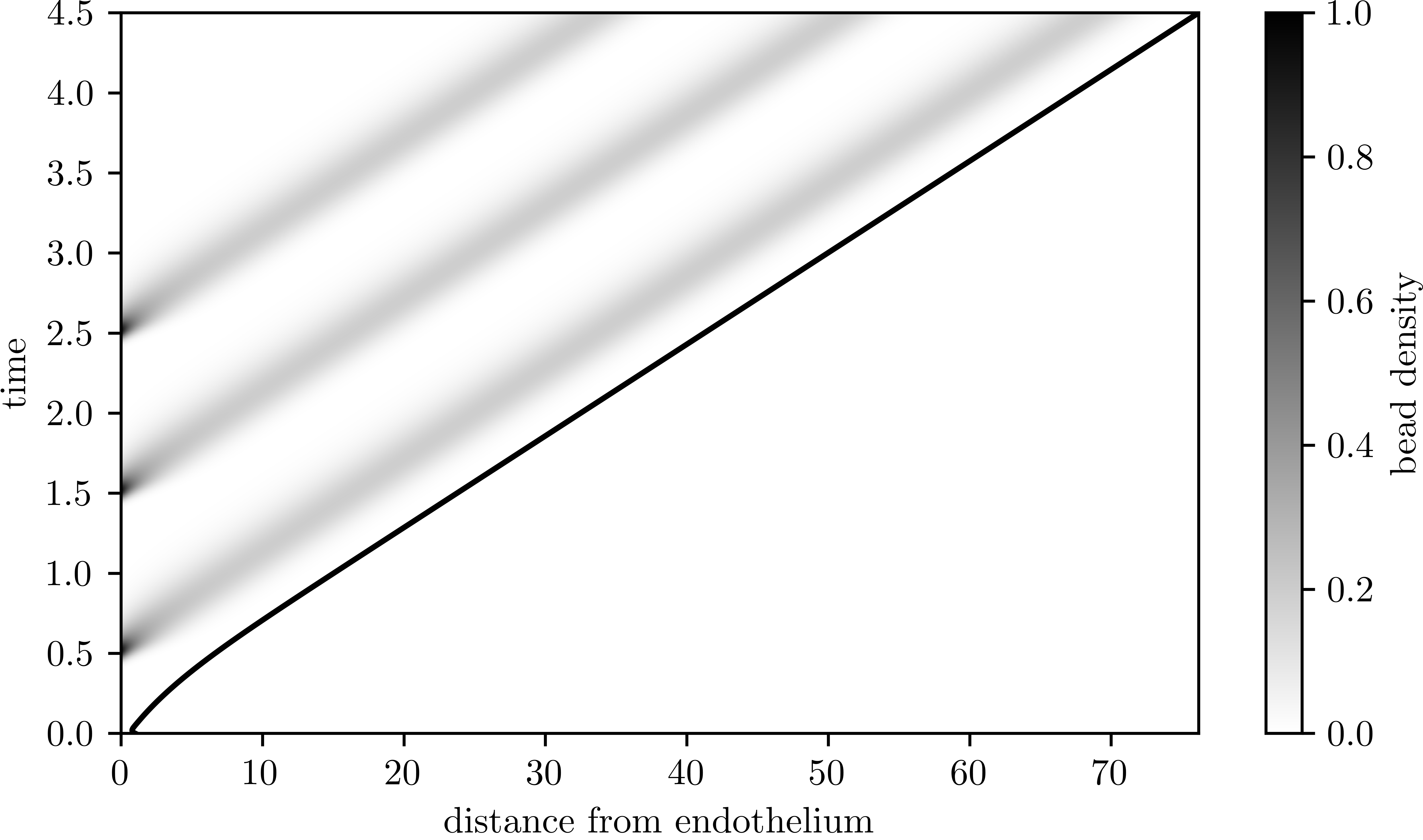}} \\ \vspace{2mm}
    (b) \includegraphics[align=t]
    {{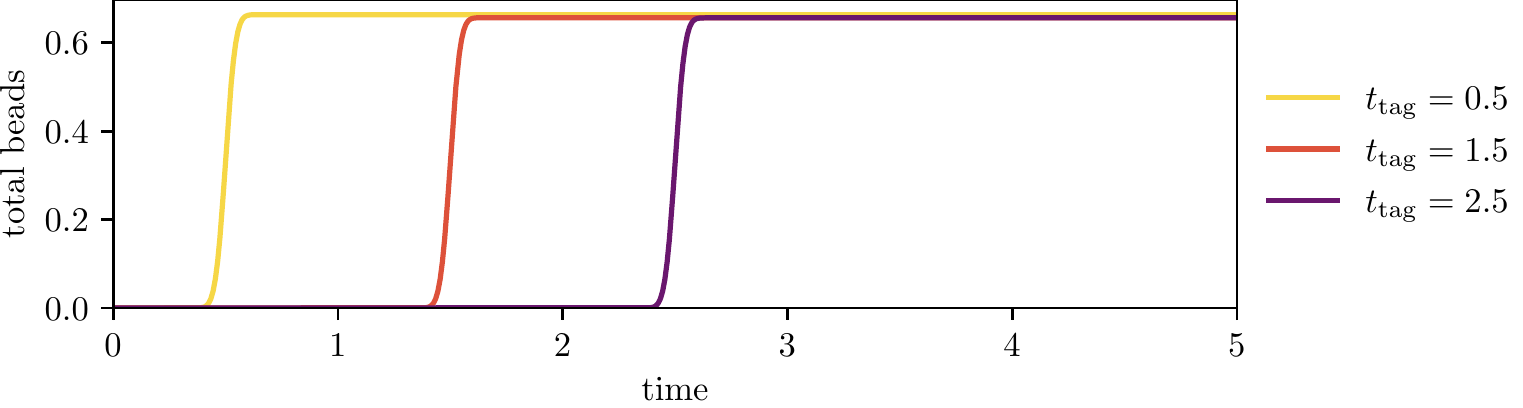}}
    \vspace{2mm}
    \caption{ Bead circulation for different bead tagging times
        for a plaque with poor efferocytosis
        ($\mu_a = 40$, $\mu_e = 30$, $\sigma_e = 40$).
        Plot (a) shows the time evolution of the total bead distribution
        $q_f(x,t) + q_c(x,t)$,
        where beads are administered multiple times at
        $t_{\text{tag}} = 0.5,\, 1.5,$ and $2.5$.
        Plot (b) compares the time evolution of
        the total bead quantity $Q(t) = \int_0^{R(t)} (q_f+q_c) \dif x$
        for the three bead dosages in (a).
    }
    \label{fig:5-pathsefflow}
\end{figure}

For plaques with poor efferocytic uptake rates
that contain mostly dead cells,
bead-tagged foam cells will die,
and the beads will remain in the plaque indefinitely.
\Cref{fig:5-pathsefflow}(a) tracks the evolution of the bead distribution
for a plaque with nonzero emigration velocity ($\sigma_e = 40$)
and poor efferocytosis
($\mu_a = 40$, $\mu_e = 80$, so $\frac{\mu_e}{\mu_a}=0.5$).
Like the high efferocytosis case,
bead-carrying cells are pushed deeper into the plaque
as new material enters through the endothelium.
Here, however, beads remain equidistant from the medial boundary
as the plaque grows.
\Cref{fig:5-pathsefflow}(a) tracks bead totals $Q(t)$ with time,
and shows that bead numbers peak soon
after the monocyte tagging time $t_{\text{tag}}$,
and continue to remain high at longer times.
This suggests that bead-carrying cells are
failing to clear from the plaque,
as no live foam cells are available to emigrate
and allow material to leave the intima.

\section{Discussion}
\label{sec:discussion}

In this work, we have constructed a new model
for early atherosclerotic plaque growth that includes
the effects of cell death on plaque development.
A key aspect of our approach is the use of a locally mass-conservative
multiphase framework with a free boundary.
This allows us to model growth of the intima
due to the accumulation of new material,
and also to model cell movement and emigration
in a manner that respects local mass conservation.

Our model finds that
dead cells are most plentiful in the deep plaque,
with a higher proportion of live macrophages 
closer to the endothelium.
This is consistent with histological studies
of early human aortic fatty streaks \citep{guyton1993transitional},
which contain macrophage-derived foam cells
mostly in their middle and closer to the endothelium.
Deposits of cholesterol crystals are observed deeper inside the intima,
which are indicative of foam cell necrosis.
Observations of more advanced plaques in humans and mice
show a similar structure,
where necrotic cores form deep in the plaque,
and live macrophages tend to localise closer to the endothelium,
albeit with a much larger necrotic core than in early fatty streaks
\citep{macneill2004focal,dickhout2011induction}.

Our model also suggests that advection is
the primary mass transport mechanism
that pushes material deeper into the plaque.
The multiphase framework we use is highly idealised,
where we include a simple mixture velocity advection term
to help enforce the no-voids condition.
This is in contrast to a force-balance equation approach
or a full continuum mechanical approach based on
porous flow theory or fluid mechanics,
where advection-type terms will arise indirectly from other assumptions.
Nevertheless,
the mixture velocity advection term here plays a similar role
to the advection term that arises in porous flow plaque models
\citep{friedman2015free,cilla2015effect},
which is typically induced by transmural pressure gradients
across the endothelium or intima .

Despite the necrotic core being a key feature of advanced plaques,
the role of dead cells in plaque development
is largely neglected by other spatial models,
and those spatial models that include cell death
tend to model it via a simple unbalanced sink term.
Ours is also the first spatially structured plaque model
to explicitly consider efferocytosis,
although \citet{ford2019lipid} consider efferocytosis 
in a lipid-structured model without spatial structure.
\citet{fok2012growth} presents a reaction-diffusion-chemotaxis model
for necrotic core formation that explicitly includes dead cells.
His model posits that the accumulation of macrophages
deep in the plaque is driven mostly by chemotaxis
towards extracellular lipid in the core,
whereas in our model,
advection is the primary means of mass transport into the deep plaque.
Fok's model also neglects the recycling
of efferocytosed material by live cells,
although it includes a term for the removal of dead cells
by other efferocytosis-competent leukocytes.

In our model, the rate of efferocytic recycling relative to cell death
is a key determinant of whether the deep plaque can support live foam cells.
Higher rates of efferocytosis relative to death
result in a higher proportion of live cells.
These results are consistent with gene knockout studies in mice,
which find that defective efferocytosis
is correlated with large plaque size
and higher numbers of dead cells \citep{kojima2017role}.
Crucially, our model undergoes a bifurcation
as the ratio of the efferocytosis and death parameters changes,
and complete foam cell death in the deep plaque is unavoidable
if the ratio falls below a threshold value.
Studies have explored the potential for using efferocytosis as a therapeutic target
to treat cardiovascular disease and cancer \citep{kojima2017role},
with strategies including
disruption of the expression of ``don't-eat-me'' signalling molecules
\citep{kojima2019cyclin},
and the use of pro-efferocytic nanoparticles \citep{flores2020pro} for example.
Our model has the unfortunate implication that
small improvements in macrophage efferocytic uptake
may not be enough to induce regression of unstable advanced plaques.
Larger improvements however may enable regression
by increasing live foam cell counts.

Foam cell emigration in our model is critical in slowing plaque growth,
and can allow plaques to stabilise if emigration is large enough.
This is consistent with experimental studies in mice
in which the CCR7 receptor is blocked.
CCR7 is involved in macrophage emigration,
and blocking the receptor leads to larger plaques
with impaired macrophage migration
\citep{luchtefeld2010chemokine}.
Intuitively, model plaques that are larger
and plaques with slower foam cell emigration
are characterised by increased retention of foam cells.
This has implications for necrotic core formation,
since foam cells that emigrate sooner
are less likely to become necrotic.
Emigration has been observed to enable steady states in other models,
including the lipid-structured model by \citet{ford2019lipid},
and in ODE models by \citet{cohen2014athero} and \citet{lui2021modelling}.
Emigration is typically neglected in existing spatial models however,
and those that do exhibit steady states
typically model emigration using unbalanced macrophage sink terms,
rather than via emigration through the domain boundaries.

Inefficient efferocytosis and high rates of cell death
inhibit foam cell emigration and plaque stabilisation
by lowering the numbers of live foam cells.
Below the threshold efferocytosis/death ratio
required for the survival of deep plaque foam cells,
emigration is effectively negligible.
High rates of macrophage death and poor rates of emigration
are both observed in plaques in hypercholesterolemic mice
\citep{feig2011hdl}.
However, it remains unclear whether the lack of live cells
specifically contributes to poor emigration rates in vivo,
or whether both are a consequence of other factors
such as excessive cholesterol loading.
Attempts to block apoptotic pathways
(using caspase inhibitors for instance)
have been inconclusive due to
the subsequent increase in necrosis
in cells with dysfunctional apoptotic pathways
\citep{martinet2011pharmacological}.

Our model provides a good foundation for developing
more detailed models of early plaque growth.
One immediately relevant question is the
effect of intracellular cholesterol,
which is known to upregulate apoptotic pathways
and disrupt normal cell function in macrophages \citep{tabas2002consequences}.
By separating the foam cell phase into two individual phases
representing macrophages and intracellular cholesterol,
we can model heterogeneity in foam cell cholesterol loads.
The intracellular cholesterol phase would inherit
the macrophages' phase velocity,
like the bead phase in this paper.
This would allow us to incorporate some of the advantages
of lipid-structured PDE models
\citep{meunier2019mathematical,ford2019lipid,watson2022lipid}
into a spatial framework,
such as cholesterol-dependent death and emigration rates.

Another important extension is the inclusion of HDL,
which has multiple atheroprotective functions
\citep{barter2004antiinflammatory,brites2017antioxidative,feig2011hdl},
including the ability to accept cholesterol from foam cells,
inhibit LDL oxidation, and stimulate macrophage emigration.
Other modelling work has shown that increasing HDL levels
can enable plaque stabilisation
\citep{thon2018quantitative,cohen2014athero,ford2019lipid,friedman2015mathematical}.
By extending our model to include HDL,
we can explicitly consider how it reduces the accumulation
of apoptotic and necrotic material.

The inclusion of a bead cargo species in our model
also provides a promising means to explore further questions
raised by \cite{williams2018limited}
on macrophage motility and retention in plaques.
Tracking individual cells is trivial in discrete cell-based models,
but presents a challenge in a continuum modelling framework,
and our bead model represents a novel method
for tracking the movement of cells within a spatial PDE model.
This approach may also prove useful in modelling other diseases
where macrophage motility is an important question,
such as in granulomas \citep{egen2008macrophage}
and tumour infiltration \citep{owen2004mathematical}.

\bibliography{refs}

\end{document}